%% file: sample-manuscript.tex
\documentclass[manuscript,screen]{acmart}
\usepackage{threeparttable, tabularx, multicol, multirow, makecell, tcolorbox, booktabs, array, amsmath, subcaption, listings, xcolor, graphicx, comment}
\usepackage[normalem]{ulem}
\AtBeginDocument{%
  }

\definecolor{YJQColor}{RGB}{0,76,153} 

\begin{document}


\title{Testing Storage-System Correctness: Challenges, Fuzzing Limitations, and AI-Augmented Opportunities}


\author{Ying Wang}
\orcid{0009-0006-3374-4403}
\email{wangying2023@ict.ac.cn}
\affiliation{%
  \institution{Institute of Computing Technology, Chinese Academy of Sciences (ICT, CAS)}
  \city{Beijing}
  \country{China}
  }

\author{Jiahui Chen}
\email{chenjiahui24@mails.ucas.ac.cn}
\affiliation{%
  \institution{Institute of Computing Technology, Chinese Academy of Sciences (ICT, CAS);  University of Chinese Academy of Sciences;
  Zhongguancun Laboratory}
  \city{Beijing}
  \country{China}
  }
  
\author{Dejun Jiang}
\email{jiangdejun@ict.ac.cn}
\affiliation{%
  \institution{Institute of Computing Technology, Chinese Academy of Sciences (ICT, CAS); University of Chinese Academy of Sciences}
  \city{Beijing}
  \country{China}
  }


\begin{abstract}
Storage systems are fundamental to modern computing infrastructures, yet ensuring their correctness remains challenging in practice. Despite decades of research on system testing, many storage-system failures---including durability, ordering, recovery, and consistency violations---remain difficult to expose systematically. This difficulty stems not primarily from insufficient testing tooling, but from intrinsic properties of storage-system execution, including nondeterministic interleavings, long-horizon state evolution, and correctness semantics that span multiple layers and execution phases.

This survey adopts a storage-centric view of system testing and organizes existing techniques according to the execution properties and failure mechanisms they target. We review a broad spectrum of approaches, ranging from concurrency testing and long-running workloads to crash-consistency analysis, hardware-level semantic validation, and distributed fault injection, and analyze their fundamental strengths and limitations. Within this framework, we examine fuzzing as an automated testing paradigm, highlighting systematic mismatches between conventional fuzzing assumptions and storage-system semantics, and discuss how recent artificial intelligence advances may complement fuzzing through state-aware and semantic guidance. Overall, this survey provides a unified perspective on storage-system correctness testing and outlines key challenges and opportunities for future semantic-aware validation.

\end{abstract}



\maketitle

\input{def}
\input{introduction}
\input{background}
\input{classical_test}
\input{fuzzing_wy}
\input{ai}
\input{conclusion}

\bibliographystyle{ACM-Reference-Format}
\bibliography{references}

\input{appendix}

\end{document}

%% file: def.tex
\newcommand{\cjh}[1]{{\color{red}#1}}

%% file: introduction.tex
\section{Introduction}
Modern computing infrastructures fundamentally rely on storage systems to preserve data integrity and sustain service reliability. Despite decades of engineering effort, storage systems continue to suffer from correctness violations that often remain undetected until they result in data corruption or large-scale service outages~\cite{bairavasundaram2008analysis, pinheiro2007failure}. As storage systems grow increasingly complex and concurrent, such failures become harder to reason about and even harder to validate in practice.

Over the years, a broad spectrum of testing techniques has been proposed to validate specific aspects of storage-system correctness~\cite{pillai2015crashconsistency, arpaci2018ostep}, including concurrency testing, long-running workload execution, crash-consistency analysis, device-level semantic validation, and distributed fault injection. While each approach is effective within a well-defined scope, these techniques are typically tailored to particular failure models, execution assumptions, or system layers. Consequently, they offer fragmented visibility into complex, long-horizon system behaviors~\cite{hydra, bairavasundaram2008analysis}, leaving many correctness failures difficult to expose systematically despite extensive testing efforts.

More fundamentally, a central challenge underlying these limitations lies in the semantic nature of storage-system correctness~\cite{nightingale2006rethink}. Unlike failures that manifest as immediate crashes, many storage-system bugs arise as latent semantic violations, whose effects emerge only under particular interleavings, after prolonged execution, or during recovery~\cite{mohan2018b3, bairavasundaram2008analysis}. Because core storage correctness properties are defined over long-horizon sequences of operations and persistent state evolution across execution phases, correctness is inherently difficult to observe, attribute, and validate using short-lived or input-local testing alone.

Given the semantic and long-horizon challenges of storage-system correctness, fuzzing naturally emerges as an appealing candidate for automated testing. Fuzzing has demonstrated broad impact across diverse software domains, including kernels~\cite{syzkaller}, compilers~\cite{grayc}, protocol implementations~\cite{restler}, and system libraries~\cite{libfuzzer}. Its success in uncovering unexpected execution paths and robustness issues suggests strong potential for storage-system testing. However, applying fuzzing to storage systems raises fundamental questions about how its underlying assumptions align with storage-specific correctness requirements. In particular, the deep statefulness and multi-phase semantics of storage systems challenge conventional fuzzing abstractions and evaluation strategies.

A growing body of survey literature has examined software testing methodologies, fuzzing techniques~\cite{zhu2022fuzzing, mallissery2023demystify}, and learning-based optimizations~\cite{qi2024surveyllmtesting, wang2024softwareLLMtesting, jiang2024whenfuzzingLLMs}, primarily from a method-centric perspective. Existing surveys typically focus on fuzzing algorithms, test-generation strategies, or artifical intelligence (AI)-assisted heuristics, without treating storage systems as a distinct testing domain characterized by long-horizon state evolution and semantic correctness requirements. As a result, there is no consolidated view that connects storage-system failure characteristics, existing testing techniques, and the roles of fuzzing and AI within a unified, storage-centric framework.

This survey addresses this gap by providing a storage-centric perspective on system testing, grounded in the semantic and stateful nature of storage-system correctness. Rather than cataloging tools or algorithms in isolation, we systematically organize existing testing techniques according to the execution properties and failure mechanisms they are designed to expose. We further examine the role of fuzzing within this landscape, identifying both its strengths and its systematic mismatches with storage-system semantics. Finally, we discuss how recent advances in AI relate to these challenges, not as generic optimizations, but as potential mechanisms for reasoning about state evolution, execution history, and semantic correctness.

Specifically, this survey makes the following contributions:
\begin{itemize}
    \item We characterize modern storage-system architectures and failure models, highlighting the semantic and execution properties that fundamentally complicate correctness testing.
    
    \item We systematize existing storage-system testing techniques by the classes of failures they target and the assumptions they rely on, clarifying their respective strengths and limitations.

    \item We examine fuzzing from a storage-centric viewpoint, decomposing the fuzzing pipeline and analyzing where existing techniques align with storage-system requirements and where they fall short.

    \item We discuss how AI-augmented techniques intersect with storage-system testing challenges, identifying key gaps between representation, decision-making, and semantic validation.
\end{itemize}

The remainder of this survey is organized as follows.
Section~\ref{sub:failure-model} reviews storage-system architectures, failure models, and correctness properties that complicate testing.
Section~\ref{sub:testing} surveys existing testing techniques for storage systems and their underlying assumptions.
Section~\ref{sub:fuzzing} summarizes existing fuzzing efforts for storage systems and highlights their limitations.
Section~\ref{sub:ai} discusses how recent AI-based approaches intersect with storage-system testing challenges, and analyzes their potential roles and limitations in addressing semantic and stateful correctness issues.
Section~\ref{sub:conclusion} concludes the survey.

%% file: background.tex
\section{Storage System Architectures and Failure Model}~\label{sub:failure-model}
Modern storage systems consist of multiple interacting layers, ranging from applications and distributed protocols to local storage engines and physical devices. The behavior of the system as a whole emerges from interactions among these layers. As a result, an architectural view provides useful context for understanding storage-system failure modes and the testing and fuzzing techniques discussed later in this survey.

The following subsections outline the major layers of contemporary storage stacks and explain how their structure shapes the space of failures encountered in practice.

\subsection{Layered Architectural Overview}

\begin{figure}[h]
  \centering
  \includegraphics[width=0.7\linewidth]{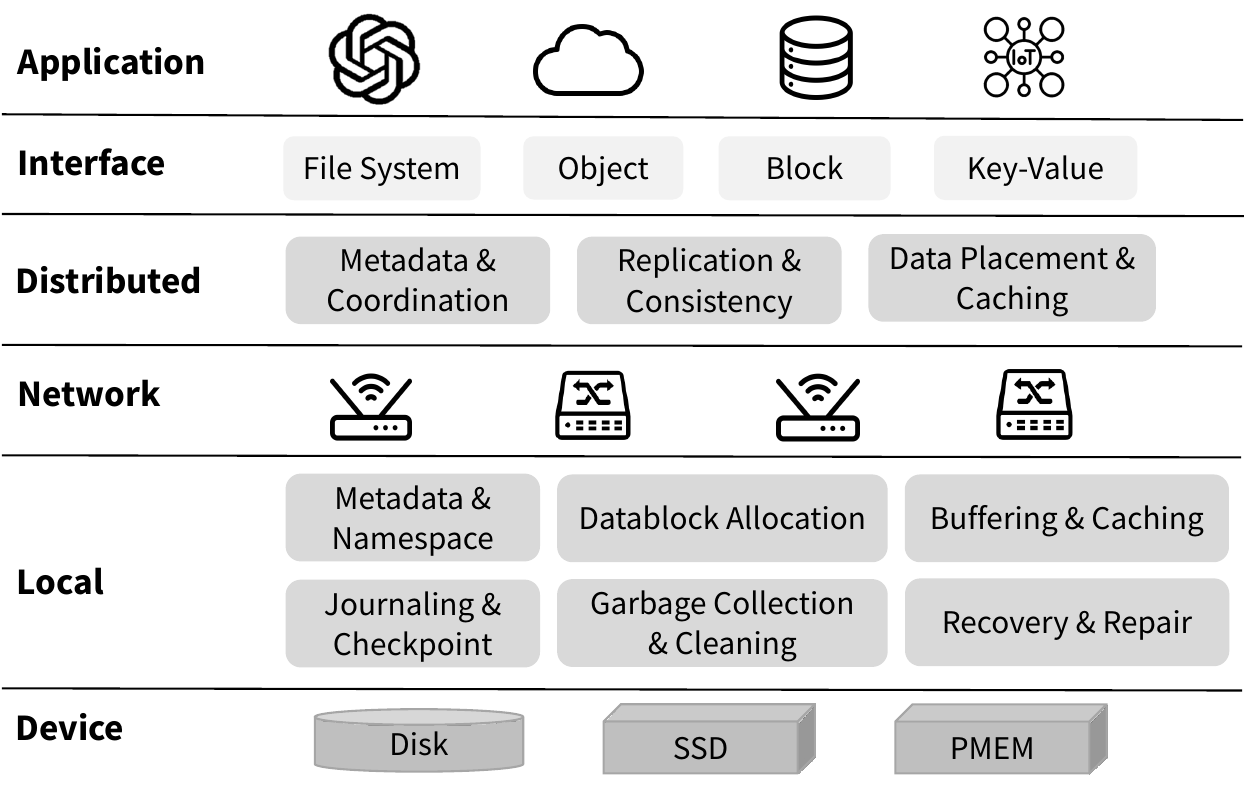}
  \caption{Multi-Layer Structure of Modern Storage Systems.}
  \label{fig:storage_architecture}
\end{figure}

Storage systems expose a variety of interface abstractions to applications, including file-based~\cite{Atlidakis16POSIX, ext4, Rodeh13BTRFS, JianXu16NOVA}, object-based~\cite{Weil06Ceph}, block-level~\cite{aws-ebs-doc, ceph-rbd-doc}, and key-value~\cite{rocksdb,Dean11LevelDB} interfaces. Despite these differences at the interface level, all such systems rely on a shared multilayer architecture beneath these abstractions. 
Figure~\ref{fig:storage_architecture} presents a representative organization of this architecture and illustrates how responsibilities are divided among layers.

At the top of the stack, the application and interface layers specify the semantic expectations that the storage system is expected to uphold.
These expectations propagate downward into the distributed layer, where replication, metadata coordination, leader election, and session management must operate correctly despite message delays, partitions, or concurrent background activities, often making violations of high-level semantics difficult to observe, attribute, and systematically test.

Beneath distributed coordination, the network and local storage layers introduce additional sources of nondeterminism. Factors such as message timing, logging and journaling behavior, allocation and cleaning policies, and recovery procedures collectively shape the system’s behavior over extended execution periods, complicating efforts to systematically explore or reason about correctness-relevant executions.
Finally, the persistence device layer exhibits latency variability, write reordering, partial persistence, and firmware-level scheduling effects, which can interact with higher layers in subtle and unexpected ways.

Although each layer has well-defined responsibilities, correctness ultimately emerges from the interaction of all layers under concurrency, failures, and long-running execution, posing persistent challenges for comprehensive correctness validation.

\subsection{Failure Model}
Building on the layered storage architecture described above, we characterize storage-system failures in terms of how architectural complexity manifests as observable system misbehavior. Although conventional bugs such as crashes or memory faults do occur, many real-world storage failures do not surface as immediate errors. Instead, they arise from violations of expected system behavior under specific timing, state-evolution, or crash-recovery conditions, making them difficult to characterize using traditional bug classifications~\cite{a_study_of_linux_file_system_evolution, neat, Dai2018TimeoutProblems, Lu2022CloudRaid, Chen2023PushButtonRainmaker, Fang2023HeteroScore}.

Rather than proposing a new taxonomy, we organize these failures into five broad, non-mutually-exclusive classes as a conceptual framework that summarizes recurring failure mechanisms observed across prior studies and systems.
The remainder of this subsection describes these failure modes in turn, focusing on their underlying semantic characteristics rather than on individual system implementations.

\subsubsection{Temporal and Ordering Failures}
These failures arise when visibility or ordering assumptions are violated under concurrency.
Foreground I/O may overlap with asynchronous background activities such as compaction or replication, and small timing variations can break prefix, durability, or atomicity guarantees without triggering a crash.
Because such violations often remain latent until later reads, they are extremely difficult to expose through sequential or coverage-based tests, highlighting the importance of testing techniques that account for timing-sensitive interleavings.

\subsubsection{State-Evolution Failures}
These failures occur when persistent structure---such as metadata, logs, and indexes---gradually drift away from their intended invariants. Errors can accumulate over long-running execution and background maintenance (e.g., compaction or scrubbing), manifesting as gradual semantic deviation. Such violations often remain latent and surface only after multiple reorganization cycles or complex coordination across components, by which point system behavior may diverge from expected semantics. This class of failures highlights how long-lived state evolution can obscure the root causes of correctness violations.

\subsubsection{Crash-Consistency and Recovery Failures}
Crashes can interrupt multi-step persistence updates, resulting in partially written data, missing flushes, or misordered log entries (e.g., metadata persisted without corresponding data blocks).
During recovery, if replay logic, metadata invariants, or cross-layer ordering constraints have been violated, the system may incorrectly reconstruct its state, leading to crash-consistency failures.
Such violations often remain latent and surface only during subsequent recovery events---such as system restarts, failover, or upgrades, highlighting the need for testing approaches that span both execution and recovery phases.

\subsubsection{Hardware- and Persistence-Model Failures}
Persistence devices introduce nondeterministic behaviors such as partial writes, latency variability, and firmware-level reordering.
These failures occur when device-level behavior deviates from the assumed persistence model, silently breaking ordering or durability guarantees even if upper layers are correct, particularly under concurrent or high-throughput workloads. They illustrate the challenges posed by mismatches between software assumptions and hardware-exposed semantics.

\subsubsection{Distributed Coordination and Replica-Consistency Failures}
Distributed systems rely on message timing, leader election, failover ordering, and coordinated version histories.
These failures occur when coordination assumptions are violated due to message delays, reordering, or independent replica progress, leading to log-prefix divergence or inconsistent visibility (e.g., replicas advancing under delayed coordination).
Such failures typically span multiple nodes and evolve over long time horizons, making them difficult to capture using isolated or single-node testing approaches.

Together, these failure modes highlight why exposing storage-system correctness violations requires testing techniques that can reason about concurrency, long-lived state, crashes, and cross-layer interactions.


%% file: classical_test.tex
\section{Practical Testing Techniques for Storage Systems} \label{sub:testing}
Testing storage systems is challenging because many correctness violations do not manifest as immediate crashes, but instead remain latent and surface only under specific conditions~\cite{filesystemsecuritywar, syncsync2024, yu2024filehijacking}. 
As a result, validating storage-system correctness requires testing techniques that go beyond crash detection and explicitly reason about timing, state evolution, and semantics across layers.  

Rather than viewing storage-system testing as a collection of unrelated tools, this section organizes existing techniques by the aspects of execution they attempt to control or observe.
Building on the failure models discussed in Section~\ref{sub:failure-model}, we analyze how different classes of testing techniques attempt to expose these failure modes in practice, emphasizing both what each class of techniques can reveal and the fundamental dimensions they leave uncovered.

\subsection{Ordering and Interleaving Testing} \label{sub:class_test_ordering}
Ordering-related violations arise under concurrent execution, when I/O operations interleave with asynchronous or background activities~\cite{Lu14Study}.
Such interleavings can break correctness guarantees, including write ordering, visibility ordering, and persistence ordering.
Because these violations occur within narrow execution windows and depend on precise execution timing, they are difficult to reliably trigger, observe, and reproduce during testing~\cite{Lu14Study, cve2015_8839, cve2024_35798}.
Techniques in this category attempt to expose correctness violations by controlling or perturbing execution order.

\subsubsection{Schedule-Controlled Concurrency Testing}
Schedule-controlled testing frameworks expose ordering-sensitive concurrency bugs by explicitly constraining and exploring thread schedules or system-call interleavings. They explore alternative executions by manipulating preemption points or enforcing specific event orders, increasing coverage of concurrency behaviors. 

Tools such as CHESS~\cite{chess} and Landslide~\cite{landslide} exemplify this approach by explicitly enumerating representative thread interleavings under bounded execution scopes. For storage systems, schedule control is most useful for components where correctness depends on subtle visibility and ordering semantics across concurrent I/O and background maintenance (e.g., log/apply paths, metadata updates, or checkpoint/compaction interactions).

However, schedule-controlled testing is largely agnostic to storage-specific semantics: it lacks guidance on which interleavings are semantically critical and must therefore bound the exploration space to remain tractable.
As a result, it primarily exposes concurrency-related ordering bugs within relatively small executions and has limited ability to capture long-horizon or multi-phase behaviors.

\subsubsection{Stress-Based Concurrency Testing}
In contrast to schedule-controlled approaches, stress-based concurrency testing relies on high load and often randomized workloads to probabilistically induce timing overlaps, interleavings, and latent race conditions.
Stress-oriented tools such as fsstress~\cite{xfstests}, dbench~\cite{dbench}, and stress-ng~\cite{stressng} generate sustained I/O pressure and large numbers of concurrent operations, increasing the likelihood of exposing concurrency-related failures.
Although these tools were originally designed as workload or stress generators rather than correctness-focused frameworks, they have historically uncovered numerous execution-ordering bugs, metadata inconsistencies, and concurrency-induced corruptions in production storage systems~\cite{a_study_of_linux_file_system_evolution}.

Importantly, stress-based concurrency testing is primarily designed to expose timing- and scheduling-dependent behaviors under high concurrency. However, because it treats execution as a black box and provides no guidance on which schedules to prioritize, coverage of ordering-sensitive behaviors remains inherently probabilistic.

\subsection{Long-Horizon and Persistent-State Testing}
Many storage-system correctness violations arise only after prolonged execution, as persistent state accumulates and evolves over time.
Such violations often depend on execution history and interactions across multiple phases, making them difficult to expose with short or isolated tests~\cite{Chen20Testing, CVE2019-19319, CVE2019-19447}. 
Techniques in this category focus on driving systems into deep, history-dependent states. 

\subsubsection{Long-Horizon Workload Testing}
Long-horizon workload testing sustains system execution over extended periods, allowing persistent state to accumulate and evolve through repeated interactions between foreground I/O and background maintenance (e.g., compaction and garbage collection).
Unlike stress-based concurrency testing, which emphasizes short-term timing effects, long-horizon workloads aim to approximate prolonged execution and state evolution.
Few existing testing frameworks explicitly target long-horizon semantic failures. Instead, prolonged workloads and endurance-style tests are commonly used in practice to approximate long-term state evolution.

In key-value and LSM-based storage engines, such prolonged workloads are commonly realized through stress and benchmark tools originally developed for performance evaluation.
RocksDB employs db\_stress and internal whitebox stress tests~\cite{rocksdb,rocksdb_stress} to drive continuous updates, deletions, and compactions, which can surface issues related to long-term version-chain evolution, metadata aging, and stale-pointer cleanup.
At the workload level, benchmarks such as YCSB~\cite{ycsb} and TPC-C~\cite{tpcc} generate sustained update-intensive executions that repeatedly exercise write amplification, background maintenance, and checkpointing across a wide range of storage engines.

For file-system-style behaviors, prolonged executions of FIO~\cite{fio} and Filebench~\cite{FileBench} (e.g., varmail and fileserver) stress allocation paths and directory mutations, surfacing fragmentation, free-space drift, and aging-related anomalies.
Complementing synthetic workloads, long-horizon trace replays such as FSL/MSR workloads~\cite{fsltraces, msrtraces} mimic multi-hour real-world usage and have revealed deep-state issues in ext4~\cite{ext4}, btrfs~\cite{Rodeh13BTRFS}, and F2FS~\cite{Lee15F2FS}.

While these workloads can reach deep internal states, they provide no explicit guidance toward semantically critical or failure-prone states, and therefore explore deep-state behaviors only probabilistically.

\subsubsection{Persistent State Validation}
Because storage systems maintain persistent state that evolves over long execution horizons, many correctness violations are not immediately observable at the point where they are introduced, but often after subsequent execution or reorganization. 
Persistent-state validation techniques examine the consistency and integrity of accumulated persistent state, focusing on end-state correctness rather than execution paths.

In practice, these techniques include both snapshot-based and online validation.
Snapshot-based validation inspects persisted state offline or periodically, comparing it across time or against reference invariants. 
RocksDB’s ldb utilities~\cite{rocksdb-github}, for example, support offline inspection of manifest evolution and SSTable layout correctness. Similarly, tools such as fsck~\cite{fsck-manpage} (ext4/btrfs) and background scrubbing mechanisms like ZFS scrub~\cite{zfs-scrub} periodically validate persisted metadata structures and allocation maps, uncovering long-term state drift. Block- and file-level integrity mechanisms such as dm-verity~\cite{dm-verity} and fs-verity~\cite{fsverity-doc} further enable snapshot-style verification by checking persisted data against Merkle-tree-based integrity metadata associated with an immutable on-disk image.

In contrast, online validation mechanisms incrementally check persistent state as it evolves during execution. Recent work proposes maintaining runtime checksums or dynamic Merkle-tree structures~(DMTs)~\cite{burke2025dynamicmerkle, liu2023dynamicmerkle} to validate storage state on the critical path. Such online approaches can detect inconsistencies earlier, but they incur continuous overhead and are therefore complementary to snapshot-based validation of accumulated persistent state.

Overall, persistent-state validation techniques are effective at detecting deep-state drift and silent corruption, but they rely on manually defined invariants and operate at a coarse semantic granularity, limiting their ability to capture transient or multi-phase violations.

\subsection{Crash-Consistency and Recovery Testing}
Crashes, power loss, and abrupt shutdowns routinely leave storage systems in partially persisted states.
Because crashes may occur at arbitrary points during execution and persistence, the space of possible post-crash states grows combinatorially with execution length, persistence ordering, and system concurrency. 
Crash-consistency testing focuses on failures where semantic violations are revealed only after execution is interrupted and recovery logic is exercised.

\subsubsection{Crash-Injection Testing}
Crash-injection testing approaches aim to explore crash-induced behaviors by executing workloads, injecting crashes at selected persistence boundaries, and validating post-recovery states against expected correctness semantics using explicit oracles or checkers.
By explicitly enumerating crash points and replaying executions under alternative persistence outcomes, these approaches enable systematic detection of recovery violations caused by partial persistence.

A representative methodology in this space is bounded black-box crash testing (B$^{3}$)~\cite{mohan2018b3}, which restricts execution to bounded workloads and explores the finite crash space induced by persistence boundaries, making systematic enumeration tractable.
CrashMonkey and ACE~\cite{crashmonkey} concretely realize the B$^{3}$ methodology for file systems. CrashMonkey provides a file-system-agnostic record-and-replay framework that logs persisted writes and constructs plausible post-crash disk states through selective reordering, while ACE improves scalability by systematically enumerating bounded workloads, thereby reducing the overall workload-crash exploration space.

Related ideas have also been applied beyond file systems. Witcher~\cite{witcher2021sosp} extends systematic crash testing to persistent-memory and key-value stores by reasoning about persistence ordering and recovery semantics at PM-aware persistence boundaries.

Despite their effectiveness in exposing crash-consistency bugs, these approaches face inherent combinatorial growth in both the crash space and the reachable system states. Each crash–replay cycle can be expensive, and coverage is ultimately constrained by crash-space explosion and replay cost.

\subsubsection{Formal Crash-Consistency Analysis}
Formal methods provide a complementary perspective to execution-based crash testing by reasoning about crash transitions and recovery semantics at the level of formal specifications or machine-checked proofs.
Frameworks such as CCFS~\cite{ccfs} and Perennial~\cite{perennial} illustrate this approach by verifying recovery invariants under explicitly modeled crash behaviors, offering stronger correctness guarantees than purely execution-driven techniques. However, these guarantees typically come at the cost of substantial manual modeling effort, and such methods do not readily scale to full production storage stacks that involve realistic concurrency, complex I/O paths, and diverse failure models.

At the extreme end of this spectrum, FSCQ~\cite{fscq} demonstrates fully machine-checked crash safety for a verified file system. While FSCQ provides strong end-to-end assurance, its verification effort remains tightly coupled to simplified system designs, making such approaches difficult to generalize to complex, evolving storage systems in practice.

\subsubsection{Specification-Guided Testing}
Several systems occupy a middle ground between pure injection-based testing and full formal verification, seeking to combine the realism of execution with selective semantic checking.
EXPLODE~\cite{yang2006explode} adapts ideas from model checking to systematically steer real storage systems toward corner cases, relying on user-provided checkers while remaining lightweight and execution-driven.
Chipmunk~\cite{chipmunk2023} extends systematic record-and-replay crash testing to persistent-memory file systems by integrating crash-state construction with domain-specific consistency checks, enabling more targeted recovery validation.

Taken together, these approaches illustrate a continuum of hybrid techniques that trade off automation, coverage, and assurance. Rather than eliminating manual effort, they often shift complexity to user-defined specifications, correctness oracles, and system-specific instrumentation, limiting scalability but enabling deeper semantic validation than purely black-box testing.

\subsection{Hardware-Software Semantic Validation}
Storage systems rely on implicit assumptions about the execution and persistence semantics provided by underlying storage devices. When device behaviors deviate from these assumptions, correctness violations may arise even if upper software layers are logically correct. 
This section illustrates that semantic gaps exist even below the software stack, reinforcing the limits of software-only testing models.

\subsubsection{Device-Level Execution Semantic Modeling}
Device-level testing techniques address hardware-exposed execution semantics by modeling internal device behaviors---such as parallelism, scheduling, and garbage collection---to study their software-visible effects on I/O ordering and timing.

Device- and firmware-level simulators such as FlashSim~\cite{flashsim} and MQSim~\cite{mqsim} exemplify this approach by enabling controlled exploration of device-internal behaviors. FlashSim models NAND-flash SSDs, capturing flash operations, Flash Translation Layer (FTL) logic, and garbage-collection effects. MQSim extends this line of work to NVMe SSDs by modeling controller parallelism, internal queueing, scheduling policies, and latency variability. By emulating device-internal execution, these simulators allow researchers to study how hardware-level behaviors propagate into higher software layers under controlled conditions, without requiring destructive testing or real hardware failures.

While these modeling-based approaches provide valuable insight into hardware–software interactions, they operate on abstracted device models and primarily capture execution semantics at the I/O interface, limiting their ability to reflect full-system behavior.

\subsubsection{Persistency and Ordering Semantic Validation}
Persistency and ordering semantic validation targets correctness issues arising from explicit persistence primitives exposed by persistent memory and modern storage interfaces (e.g., store, flush, and fence). Because durability, atomicity, and crash consistency depend on subtle ordering relationships among these primitives that are not captured by conventional crash-based testing, a class of tools focuses on systematically exploring persistence-ordering behaviors under the hardware persistence model. 

Tools such as pmreorder~\cite{pmreorder}, XFDetector~\cite{xfdetector}, and PMTest~\cite{pmtest} exercise alternative orderings of stores, flushes, and persistence events to validate assumed correctness guarantees.
pmreorder exposes ordering-related durability violations by systematically reordering persist operations.
XFDetector identifies missing, misplaced, or incorrectly ordered flushes that silently break persistence semantics.
PMTest injects crashes as a mechanism to perturb persistency boundaries, uncovering atomicity and recovery violations that are invisible under traditional block-based crash testing.

Together, these tools provide systematic coverage of persistence-ordering semantics at the hardware–software interface. However, they primarily reason about low-level persistency behaviors and offer limited visibility into higher-level system interactions or long-horizon semantic effects.

\subsection{Distributed-Consistency Testing}\label{sub:class_test_distribute}
Distributed storage systems must maintain correctness in the presence of replication, coordination, and failures across multiple nodes.
Message delays, partitions, and partial failures can cause replicas to diverge and violate cross-replica semantic guarantees such as visibility, ordering, or durability, requiring testing to explicitly exercise fault scenarios and check semantic consistency across replicas.
Testing techniques in this category explicitly target distributed nondeterminism across replicas.

\subsubsection{Chaos Testing with Network/Process Faults}
A common approach to testing distributed storage systems is to inject network- and process-level faults during execution, as exemplified by chaos testing.
Jepsen-style testing~\cite{jepsen} injects faults such as network partitions, message delays, and process pauses during execution, and has uncovered numerous correctness violations in widely deployed storage and database systems.

NEAT~\cite{neat} explores richer network behaviors, including partial partitions and asymmetric reachability, showing that simplistic partition assumptions can mask protocol vulnerabilities. Other studies~\cite{timeoutstudy2018, iaso} further show that latency variation and fail-slow behavior can trigger incorrect leader elections, replication stalls, and faulty failover paths in production clusters.

Overall, chaos testing has proven effective at uncovering distributed failures under adverse network and process conditions. However, its fault injection is largely ad hoc and coverage-unaware, leaving much of the distributed fault and execution space unexplored and motivating complementary, more systematic approaches.

\subsubsection{Systematic Exploration of Faults and Nondeterminism}
In contrast to chaos testing, which relies on randomized or experience-driven fault injection, systematic exploration techniques treat faults and message-delivery choices as explicit dimensions of execution.

Molly~\cite{molly} identifies causally relevant faults using lineage information and systematically constructs targeted multi-fault scenarios that are unlikely to be exercised by random injection.
SAMC~\cite{samc} models message delivery choices as schedulable events and uses DPOR-style reductions to enumerate representative delivery interleavings, exposing coordination bugs that are difficult to trigger with conventional chaos testing.

Together, these techniques illustrate the benefits of structured exploration over randomized fault injection, but they require additional system modeling or tracing support and face scalability challenges as fault, timing, and state spaces grow.

\subsubsection{Multi-Replica Consistency Testing}
Multi-replica storage systems must preserve semantic consistency across replicas despite failures and concurrency. Testing such systems therefore centers on detecting replica divergence, either in client-observed operation ordering or in internal replica state.

One line of work approaches this problem from the client’s perspective. Jepsen-style testing~\cite{jepsen} reconstructs execution histories and checks consistency properties such as linearizability and related client-observable consistency guarantees, and has uncovered numerous violations in production systems whose replication protocols appear correct in isolation.

A complementary line of work focuses on validation at the protocol level. Replication protocols such as Raft~\cite{ongaro2014raft} and Paxos~\cite{lamport1998paxos} define safety properties (e.g., log-prefix consistency and ordered commits) that are not fully captured by client-visible behaviors alone. Runtime refinement checkers such as RPRC~\cite{rprc2025} bring this perspective to production systems by monitoring live executions in Etcd~\cite{etcd}, Redis-Raft~\cite{redis-raft}, and ZooKeeper~\cite{hunt2010zookeeper}, and comparing observed behaviors against formal specifications.

Beyond runtime monitoring, Mocket~\cite{mocket} performs counterexample-driven, systematic exploration by constructing adversarial message schedules and execution scenarios that are likely to violate protocol invariants, exposing rare interleavings that random scheduling often misses.

Overall, these approaches are effective at detecting replica-level semantic violations, but they require explicit specifications and checkers and tend to validate externally observable outcomes or committed protocol events. As a result, they can miss transient, multi-phase behaviors---such as partially applied operations or temporary replica divergence---that arise during execution and disappear after convergence.

\begin{table*}[!htb]
\centering
\caption{Existing testing techniques for storage systems.}
\label{tab:classical_testing}
\small 

\begin{tabularx}{\textwidth}{>{\raggedright\arraybackslash}p{2.8cm} >{\raggedright\arraybackslash}p{2.5cm} >{\raggedright\arraybackslash}X >{\raggedright\arraybackslash}p{2.6cm}}
\toprule
\textbf{Category} & \textbf{Technique} & \textbf{Representative Tools} & \textbf{Key Limitations} \\ \midrule

\multirow[t]{2}{=}{Ordering \& Interleaving} 
& Schedule Control & Chess~\cite{chess}, Landslide~\cite{landslide} & No semantic guidance \\ \cmidrule{2-4}
& Stress \& Execution & fsstress~\cite{xfstests}, dbench~\cite{dbench}, stress-ng~\cite{stressng} & Probabilistic coverage \\ \midrule

\multirow[t]{2}{=}{Long-Horizon \& Persistent-State} 
& Workload Endurance & db\_stress\cite{rocksdb}, YCSB~\cite{ycsb}, TPC-C~\cite{tpcc}, FIO~\cite{fio}, Filebench~\cite{FileBench}, FSL/MSR traces\cite{fsltraces, msrtraces} & Probabilistic deep-state reachability \\ \cmidrule{2-4}
& State Validation & ldb~\cite{rocksdb-github}, fs-verity~\cite{fsverity-doc}, fsck~\cite{fsck-manpage}, ZFS scrub~\cite{zfs-scrub}, dm-verity~\cite{dm-verity} & Coarse-grained invariants \\ \midrule

\multirow[t]{3}{=}{Crash Consistency \& Recovery} 
& Crash-Injection Testing & CrashMonkey, ACE~\cite{crashmonkey}, $B^{3}$~\cite{mohan2018b3}, Witcher~\cite{witcher2021sosp} & State-space explosion \\ \cmidrule{2-4}
& Formal Analysis & CCFS~\cite{ccfs}, FSCQ~\cite{fscq}, Perennial~\cite{perennial} & High modeling cost \\ \cmidrule{2-4}
& Specification-Guided Testing & EXPLODE~\cite{yang2006explode}, Chipmunk~\cite{chipmunk2023} & Partial specifications \\ \midrule

\multirow[t]{2}{=}{Hardware-Software Semantics} 
& Device Semantic Modeling & FlashSim~\cite{flashsim}, MQSim~\cite{mqsim} & Abstracted device model \\ \cmidrule{2-4}
& Persistency Validation & pmreorder~\cite{pmreorder}, XFDetector~\cite{xfdetector}, PMTest~\cite{pmtest} & Interface-level only \\ \midrule

\multirow[t]{3}{=}{Distributed Consistency} 
& Chaos Testing & Jepsen~\cite{jepsen}, Neat~\cite{neat}, IASO~\cite{iaso} & Ad hoc fault coverage \\ \cmidrule{2-4}
& Systematic Exploration & Molly~\cite{molly}, SAMC~\cite{samc} & Scalability limits \\ \cmidrule{2-4}
& Replica Consistency Checking & Jepsen~\cite{jepsen}, RPRC~\cite{rprc2025}, Mocket~\cite{mocket} & High oracle cost \\ 
\bottomrule
\end{tabularx}
\end{table*}
Table~\ref{tab:classical_testing} summarizes the existing testing techniques discussed in the preceding sections (Sections~\ref{sub:class_test_ordering}-\ref{sub:class_test_distribute}), organizing them by failure dimension, representative tools, and their fundamental limitations. Despite their diversity, these techniques share common fundamental limitations that stem not from tooling immaturity, but from fundamental properties of storage-system execution. In the next section, we analyze these root causes and explain why storage systems remain fundamentally difficult to test despite decades of testing research.

\subsection{Root Causes of Storage-System Testing Complexity}~\label{test_challenges}
While the failure model in Section~\ref{sub:failure-model} characterizes what can go wrong in storage systems and the preceding sections survey how existing testing techniques attempt to expose these failures, many storage-system failures nevertheless remain difficult to test in practice.
This section identifies a set of underlying execution characteristics that fundamentally limit testing automation and coverage, independent of any specific failure mode or testing tool.

At a high level, storage systems operate under nondeterministic interleavings, evolve long-lived internal state, and enforce semantic guarantees that span layers and execution phases. As a result, correctness violations often stem from timing, history, and cross-layer interactions, rather than from malformed inputs or isolated execution errors.

\begin{figure}[H]
	\begin{center}
		\includegraphics[width=0.8\linewidth]{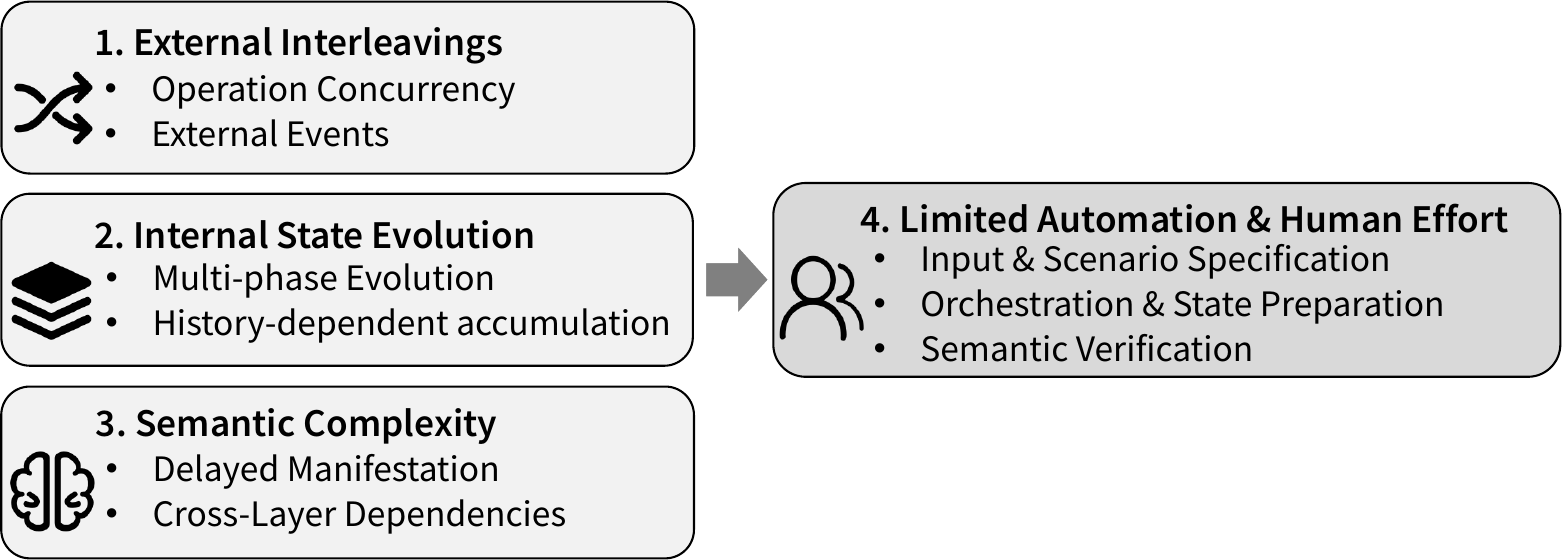} 
	\end{center}
	\caption{Four fundamental dimensions of storage-system testing complexity. }\label{fig:storage_testing_problem}
\end{figure}

Figure~\ref{fig:storage_testing_problem} summarizes four fundamental dimensions of storage-system testing complexity that underlie these challenges and explain why many correctness bugs remain difficult to trigger, observe, and validate systematically.

\subsubsection{External Interleavings}
External interleavings capture nondeterministic overlaps between normal storage operations and events originating outside the system’s control. 
Storage operations such as reads, writes, and flushes execute concurrently across threads and nodes~\cite{Weil06Ceph, ext4}, producing fine-grained interleavings that depend on scheduler timing and system load. Such execution-driven nondeterminism can lead to divergent outcomes that are difficult to enumerate or reproduce.

In addition to concurrency-induced interleavings, storage systems must tolerate external failure events such as crashes, power loss, device errors, and network partitions~\cite{Ghemawat03Google, Shvachko10Hadoop, Calder11Windows}. These events may occur at arbitrary execution points and fundamentally alter the outcome of otherwise identical operation sequences. When failures interleave unpredictably with ongoing operations, the resulting schedule space grows combinatorially, making exhaustive exploration infeasible.

\subsubsection{Internal State Evolution}
Storage systems transition through multiple internal states due to interactions between foreground operations and asynchronous background activities, causing execution to progress through multiple internal phases. Examples of such interactions include foreground I/O operations interleaving with background activities such as compaction and log replay~\cite{Verbitski17Amazon, Zhai2020CheckBeforeYouChange}.
Consequently, executions that differ only within narrow timing windows may traverse different state-transition paths, leading to a combinatorial explosion of reachable states that is difficult to enumerate or reproduce during testing.

Storage state is inherently history-dependent. Persistent structures---such as logs, metadata versions, allocation layouts, and replica histories---accumulate over time and encode the effects of prior executions, failures, and recoveries. Consequently, the same external workload can lead to different behaviors depending on when it is applied, and many critical states emerge only after prolonged execution or repeated recovery~\cite{a_study_of_linux_file_system_evolution, crashmonkey}. This long-term dependence on execution history fundamentally limits the effectiveness of short or isolated tests, which rarely reach the internal states where subtle correctness violations manifest.

\subsubsection{Semantic Complexity}
Beyond nondeterministic execution and state evolution, a fundamental source of testing difficulty lies in the semantic nature of storage-system correctness itself. 
Storage-system correctness is defined in terms of higher-level semantic expectations that relate external operations, execution history, and persistent state over time.
As a result, semantic violations do not necessarily manifest at the point where they are introduced~\cite{hydra, Wu2025Fawkes}.
Instead, their effects may remain latent and become observable in later execution phases, such as crash recovery, log replay, or subsequent reads. This temporal separation between cause and observable effect complicates both detection and attribution during testing.

More fundamentally, storage-system semantics are defined over end-to-end behaviors that span multiple layers and execution phases~\cite{hydra, LeBlanc2022FlyTrap, Rebello2020FsyncFailures}. Tests rarely consist of isolated operations; rather, they exercise sequences whose meaning depends on ordering, timing, and persistence boundaries. For example, a logical write may be logged, later propagated through background write-back, and eventually affected by device-internal reorganization; observing any single step in isolation is insufficient to determine whether the intended semantic guarantee has been satisfied. 

Consequently, assessing semantic correctness requires correlating information distributed across layers and time, forcing tests to decide when and under what execution context correctness should be evaluated. This substantially increases the difficulty of constructing automated tests and interpreting their outcomes.

\subsubsection{Implications for Automation and Human Effort}
In practice, testing storage systems continues to require substantial human involvement.
The challenges discussed above---external interleavings, long-horizon internal state evolution, and cross-layer semantic complexity---collectively make it difficult to structure storage-system testing in a fully systematic manner.
As a result, many critical testing decisions, including which executions to explore, which internal states to reach, and when correctness should be evaluated, remain difficult to derive in a principled way and therefore continue to rely on human expertise.

Storage systems expose rich interfaces~\cite{Atlidakis16POSIX} and configuration spaces~\cite{Sehgal15empirical}, and their behavior is further shaped by external events such as crashes, power loss, and device errors~\cite{Shvachko10Hadoop}. Together, these factors give rise to a combinatorial space of possible executions. 
While automated tools can generate large volumes of operations or inject generic faults, they lack the semantic context needed to prioritize operation sequences, ordering dependencies, or failure scenarios that are most likely to violate system guarantees. Consequently, the construction of effective workloads and fault models remains largely manual.

Moreover, many storage bugs manifest only after the system reaches specific internal states shaped by long execution histories, background activities, or recovery cycles~\cite{hydra, Wu2025Fawkes, crashmonkey}. Reaching such states requires deliberately steering execution through particular sequences and timing conditions, rather than simply increasing test volume. Existing tools lack models of internal state evolution and therefore cannot reliably identify semantically interesting phases or guide execution toward them.

Finally, determining whether a storage system behaves correctly requires interpreting outcomes in terms of semantic guarantees rather than detecting crashes alone. Many failures appear as subtle violations of durability, ordering, or consistency and cannot be captured by simple assertions. Defining correctness criteria, specifying invariants over persistent state, and implementing semantic oracles remain highly system-specific tasks, dependent on execution history, timing, and cross-layer interactions. As a result, semantic validation continues to demand substantial manual effort.

In summary, the continued reliance on human effort reflects fundamental limits in the degree of automation achievable by existing storage-system testing techniques.
Nondeterministic interleavings, long-term state evolution, and cross-layer semantics jointly constrain how systematically executions can be generated, guided, and validated, highlighting the need for testing approaches that can better incorporate semantic awareness and state-sensitive guidance.

\subsection{Summary}
Existing testing techniques for storage systems span a wide range of approaches, including concurrency testing, long-horizon workload execution, persistent-state validation, crash-consistency analysis, hardware-semantic testing, and distributed fault injection.
While effective at exposing specific classes of failures, these techniques primarily address isolated dimensions of storage-system execution and provide limited support for systematically coordinating execution control, state exploration, and semantic validation.

This fragmentation motivates the need for testing approaches that can adapt execution based on observed system state and semantic feedback---an issue we revisit when examining fuzzing and its extensions in later sections.

%% file: fuzzing_wy.tex
\section{Foundations of Fuzzing and Advances Toward Storage Systems}~\label{sub:fuzzing}
The limitations of existing storage-system testing techniques arise from fundamental properties of storage systems themselves. Highly concurrent execution, long-lived and evolving persistent state, and correctness semantics that span multiple layers make storage systems difficult to exercise and validate systematically, often forcing testing workflows to rely on substantial human involvement.

Against this backdrop, fuzzing has attracted growing interest as a potential path toward reducing manual effort in system testing.
By leveraging algorithmic input generation---typically through stochastic mutation or grammar-based synthesis---and using feedback signals to guide exploration, fuzzing has dramatically reduced testing effort in kernels~\cite{syzkaller}, libraries~\cite{libfuzzer}, compilers~\cite{csmith}, and networked services~\cite{restler}.

However, whether these successes translate to storage systems is not obvious.
Despite its effectiveness in many domains, it remains unclear whether fuzzing is well suited to the semantic and long-horizon correctness properties inherent to storage systems. In particular, it is not obvious whether fuzzing can meaningfully reduce the testing burdens observed in existing storage-system testing, or whether it encounters similar limitations when confronted with concurrency, persistent state, and cross-layer correctness requirements.

This section examines fuzzing from a storage-centric perspective. Rather than cataloging fuzzing techniques in isolation, we ask a more targeted question: to what extent can fuzzing address the fundamental storage-system testing challenges identified earlier?
To answer this question, we first review the core fuzzing pipeline and recent advances at each stage, and then analyze how these mechanisms interact with storage-specific properties such as concurrency, persistent state evolution, and cross-layer correctness semantics.

\subsection{Fuzzing Capabilities for Storage Systems}
We focus on fuzzing in this section because it represents one of the most widely adopted approaches to automated software testing. Understanding what fuzzing can—and cannot—systematically explore is essential to assessing its practical value for storage-system correctness validation.

\subsubsection{The Fuzzing Pipeline}
\begin{figure*}[htbp]
    \centering
    \includegraphics[width=\textwidth]{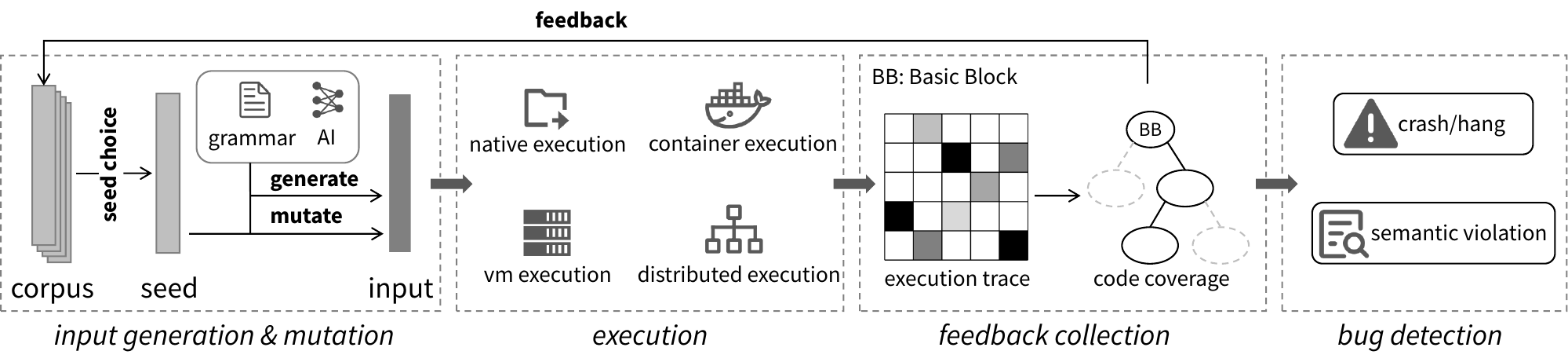}
    \caption{Modern Fuzzing Pipeline.}
    \label{fig:fuzzing_pipeline}
\end{figure*}

A fuzzing campaign typically follows a cyclic, feedback-driven workflow composed of four major stages, as illustrated in Fig.~\ref{fig:fuzzing_pipeline}:
\begin{itemize}
\item \textbf{Input Generation and Mutation:} Producing test cases by either synthesizing new inputs from scratch based on models or grammars or transforming seeds selected from a corpus---initially provided by users or accumulated during the process. 
\item \textbf{Execution:} Running each input on the target system through native execution or inside an isolated environment such as a virtual machine, container, or emulator.
\item \textbf{Feedback Collection:} Monitoring execution signals to evaluate input quality, guiding seed selection, corpus updates, and mutation strategies.
\item \textbf{Bug Detection:} Identifying anomalies such as crashes, assertion failures, or semantic inconsistencies.
\end{itemize}
This iterative loop continuously refines the input corpus, improving both coverage and depth of exploration. 
While concrete realizations of each stage may vary widely---ranging from grammar- or learning-based generation to native, virtualized, or distributed execution---the pipeline abstraction provides a common framework for understanding and analyzing modern fuzzing systems.

\subsubsection{Input Generation and Mutation}~\label{sub:input_generation_and_mutation}
A core strength of modern fuzzers lies in their ability to automatically generate and mutate large volumes of inputs, enabling broad exploration with relatively limited manual effort.
Depending on the target system and interface, fuzzing inputs can be defined at different levels of abstraction, ranging from individual input parameters, to sequences of operations, and even to system state.

\textbf{Parameter-level input mutation.}
Parameter-level mutation constitutes a fundamental building block of modern fuzzing. At this level, fuzzers modify individual input parameters---ranging from bit-level changes to perturbations of structured fields---to explore program behaviors efficiently and at scale.

For storage systems, many parameters are correctness-critical because they influence persistent-state layout, alignment constraints, allocation behavior, and recovery outcomes. Such parameters determine not only which regions of persistent state are accessed or modified, but also how updates interact with allocation policies and recovery mechanisms. Although parameter-level mutation remains essential, existing fuzzing approaches---including widely used system-call fuzzers~\cite{syzkaller} and file-system fuzzing frameworks~\cite{hydra, mock}---largely treat parameters as syntactic objects, with limited awareness of storage-specific semantic roles, making parameter-level mutation necessary but insufficient for exercising deeper storage semantics.

\textbf{Sequence-level input generation. }
In stateful systems, the logic and correctness of a single operation are often contingent upon the accumulated state produced by the history of preceding calls. Modern fuzzers therefore extend input generation beyond individual parameters to sequences of operations, enabling exploration of ordering effects, inter-operation dependencies, and higher-level behavioral patterns such as resource lifecycles, protocol phases, and invariant transitions. As a result, sequence-level generation has become a standard mechanism for exercising stateful behaviors in complex systems.

At this level, fuzzing typically needs to satisfy two forms of correctness.
On the structural side, operation sequences are often constrained by complex syntax, data types, and protocol formats. Grammar-aware fuzzers address this challenge by relying on explicitly specified grammars or interface descriptions, as exemplified by tools such as Nautilus~\cite{nautilus} and Grammar-Mutator~\cite{grammar-mutator}.
Syzkaller~\cite{syzkaller} operationalizes this approach through its domain-specific language, which encodes system-call interfaces, argument types, and protocol formats to generate well-formed invocations.

Beyond structural correctness, effective sequence generation must also preserve semantic correctness.
In existing fuzzing systems, semantic correctness is often approximated through operation-level dependencies, such as producer-consumer relationships or basic resource lifecycle constraints.
For example, in POSIX-style interfaces~\cite{Atlidakis16POSIX}, the return value of one system call may serve as an input parameter to a subsequent call, requiring fuzzers to respect basic dependency constraints to generate valid sequences.
To support such semantics, widely used fuzzers employ dependency tracking mechanisms, such as choice table or operation relation table, as seen in systems like Syzkaller~\cite{syzkaller} and Horcrux~\cite{horcrux}.

However, these dependency-based models primarily capture local and short-range semantic relationships among operations.
While they are effective at maintaining basic sequence validity, they typically cover only a limited subset of the semantic space required by storage systems.
Several systems attempt to infer richer dependencies automatically.
For example, Healer~\cite{healer} learns influence relations among system calls to guide mutation toward semantically meaningful sequences, and MOCK~\cite{mock} leverages neural language models to capture contextual dependencies from execution traces.
Nevertheless, even these approaches generally remain limited when confronted with the broader semantic space induced by concurrency, distributed coordination, protocol state, and long-lived persistent behavior that characterize storage systems in practice.

\textbf{State-level input extension.}
As fuzzing is increasingly applied to stateful systems, many behaviors of interest depend not only on the sequence of operations, but also on the accumulated system state produced by prior execution and external events.
In such settings, starting each fuzzing run from a clean state often requires replaying long and specific prefixes of operations and execution conditions to reach meaningful states, increasing execution cost and limiting the diversity of states that can be explored within a fixed fuzzing budget.

Storage systems amplify this challenge because many correctness violations emerge only after prolonged execution, background maintenance, or repeated recovery cycles have reshaped long-lived persistent state and metadata~\cite{hydra, janus}.
To mitigate the cost of repeatedly reaching deep and history-dependent states through long execution prefixes, recent work extends the notion of fuzzing input by explicitly incorporating system state into the input space, allowing fuzzing to start from valid and semantically meaningful system contexts without replaying full execution histories.
For local file systems, systems such as JANUS~\cite{janus} and HYDRA~\cite{hydra} treat the on-disk image as part of the fuzzing input. The image represents a non-empty persistent state that would normally require substantial prior execution to reach, and is mutated to expose inconsistencies that arise under specific disk layouts or accumulated metadata conditions.

More broadly, in distributed storage systems, system state is shaped not only by persistent data but also by fault and coordination context. Approaches such as MONARCH~\cite{monarch} and Hydra~\cite{hydra} treat faults---including crashes, network partitions, and message reordering---as additional state dimensions that shape system evolution in ways not captured by operation sequences alone, and interact with operation sequences over time.
Some systems further combine these dimensions by generating concurrent operation sequences across multiple nodes, effectively fuzzing both execution and state evolution together.

In practice, incorporating persistent state and failures into the fuzzing input space requires substantial system-specific modeling, and the resulting state-space explosion of state and execution histories limits both scalability and the degree of automation achievable in practice.

Taken together, these approaches illustrate a progressive expansion of the fuzzing input space---from individual parameters, to operation sequences, and finally to persistent state and failures.
While this expansion enables deeper exploration of storage-system behaviors, it also introduces increasing dependence on system-specific structure and semantic assumptions, which in practice complicates input generation and limits portability across diverse storage systems and execution scenarios.

\subsubsection{Execution Mechanism}
Execution in fuzzing is designed to minimize manual intervention while enabling repeated and systematic exploration of program behaviors.
At its core, the execution abstraction assumes that individual test cases can be replayed from a well-defined initial state and executed in isolation, allowing observed behaviors to be attributed to specific inputs and analyzed consistently across runs.
In practice, such isolation is primarily approximated by resetting execution state between test cases.

In general-purpose fuzzing of stateless or short-lived programs, reproducibility is typically achieved by restarting the program for each test case, which naturally resets execution to a clean initial state. This assumption breaks down for stateful systems---such as operating systems and storage systems---where executions may leave persistent or long-lived effects that influence subsequent runs. To approximate reproducible and isolated execution under these conditions, fuzzing frameworks execute tests within environments that can be reset to a known baseline, thereby limiting cross-test interference.

Syzkaller~\cite{syzkaller}, for example, runs the operating system inside a virtual machine and periodically reboots the virtual machine to restore system state.
However, because large numbers of test cases are executed between reboots, operating-system aging effects accumulate, reducing reset fidelity and making it difficult to precisely reproduce failures that depend on subtle timing, ordering, or long-lived state interactions.
HYDRA~\cite{hydra} and JANUS~\cite{janus} adopt finer-grained reset mechanisms by executing each test sequence within a clean or snapshot-based environment using the Linux Kernel Library (LKL). By forking from a pristine snapshot, these systems prevent persistent-state mutations from leaking across executions while significantly reducing reset overhead compared to full virtual-machine reboots.

Despite these engineering refinements, reset-based execution provides only an approximate form of isolation for storage systems.
In particular, file systems and distributed storage systems often involve executions that span multiple components or nodes, with coordinated metadata, replicated state, and background services participating in a single logical operation.
Resetting execution in such settings requires restoring not only local process or kernel state, but also distributed metadata and coordination context, substantially increasing reset complexity and cost.
As a result, reset-based isolation becomes increasingly coarse-grained and expensive, reflecting a fundamental mismatch between the execution abstraction assumed by fuzzing and the execution realities of stateful and distributed storage systems.

\subsubsection{Feedback Mechanism}
Fuzzing relies on feedback to identify “interesting” inputs and guide exploration. 
Accordingly, modern fuzzing systems typically adopt coverage-based guidance, sometimes augmented with target- or domain-specific feedback signals to steer exploration toward particular behaviors of interest.

\textbf{Coverage-based Feedback}
Control-flow coverage has long been the dominant feedback signal in fuzzing, as adopted by widely used systems such as AFL~\cite{afl}, libFuzzer~\cite{libfuzzer}, and their derivatives.
In this paradigm, execution progress is measured by whether an input exercises previously unseen basic blocks or control-flow edges, providing a lightweight and generic notion of novelty that scales well across programs.

In these systems, critical failures often stem not from unexplored control-flow paths, but from different combination and ordering of operations that may exercise already-covered code in semantically distinct contexts~\cite{janus, hydra, crashmonkey}.
As a result, control-flow coverage can provide an incomplete proxy for semantic progress, and severe correctness violations—such as ordering errors, durability breaches, or recovery inconsistencies—can occur even when execution remains within already-covered code.

Coverage-guided fuzzers therefore rely on control-flow novelty as a proxy for semantic progress. Under this approximation, executions that are semantically distinct but control-flow equivalent receive little additional attention, leading fuzzers to distribute testing effort uniformly across code regions. This mismatch reflects a structural limitation of coverage as a progress metric, rather than a deficiency of specific fuzzing implementations.

\textbf{Semantic-aware Feedback Mechanisms}
Beyond coverage-based feedback, several systems enhance fuzzing guidance with semantic awareness to better target storage-specific bugs.
Rather than treating control-flow novelty as progress, these approaches attempt to approximate semantic progress.

For instance, KRACE~\cite{krace} targets concurrency bugs in file systems by introducing alias coverage, which captures cross-thread memory-access overlaps instead of basic-block execution.
Horcrux~\cite{horcrux} detects metadata inconsistencies in distributed file systems by measuring the duration of inconsistency across replicas, rather than relying on control-flow signals.
More generally, systems such as Actor~\cite{actor} abstract execution into higher-level actions and protocol stages, guiding exploration based on meaningful state transitions rather than low-level code coverage.

While such semantic-aware feedback improves precision, it introduces new constraints on automation and generality.
Unlike coverage, semantic signals typically require substantial system-specific modeling and manual design effort, reducing their portability across storage systems.

In summary, feedback design plays a central role in determining what fuzzing explores and what it overlooks.
While coverage-based feedback offers scalability and generality, it fails to capture semantic progress in stateful storage systems.
Semantic-aware feedback can improve precision, but does so at the cost of increased system-specific effort and reduced automation, exposing a fundamental tradeoff in storage-system fuzzing.

\subsubsection{Bug Detection}
A practical strength of fuzzing is its ability to automatically detect many common bug classes without requiring explicit specifications or manual interpretation.
In practice, many fuzzing frameworks rely on readily observable execution failures---such as crashes, hangs, assertion violations, and sanitizer reports---as default bug oracles.
However, such oracles primarily capture failures that manifest as immediate execution anomalies, leaving deeper, state-dependent correctness issues unaddressed.

\textbf{Check-based semantic validation. }
Many storage bugs manifest as semantic violations rather than immediate crashes, with effects that are delayed, state-dependent, or distributed across execution phases and software layers.
Examples include metadata inconsistencies, incorrect recovery outcomes, replica divergence, and violations of protocol-level guarantees~\cite{hydra, horcrux, jepsen}. 
Because such failures do not terminate execution, they remain invisible to crash- or sanitizer-based oracles.

To detect such bugs, recent storage-system fuzzers increasingly integrate explicit semantic checkers that encode domain-specific correctness invariants. 
For example, KRACE~\cite{krace} embeds a data-race detector to uncover concurrency bugs in file systems, while Horcrux~\cite{horcrux} introduces consistency monitors to detect replica divergence beyond acceptable windows.
Hydra~\cite{hydra} provides an extensible framework that integrates multiple semantic checkers---including crash-consistency and POSIX-violation detectors---and MONARCH~\cite{monarch} extends this checker-based design to distributed storage semantics.

Crash- and sanitizer-based oracles scale well but are insufficient for storage systems, where many failures manifest as semantic violations rather than immediate execution anomalies. Checker-based validation improves semantic coverage, but its applicability is fundamentally constrained by the large and evolving space of storage-system semantics, persistent states, and execution scenarios. As a result, practical storage-system fuzzing often depends on a limited number of carefully designed checkers, trading generality and automation for focused semantic validation.

\subsection{Limitations for Storage Context}
The preceding analysis shows that, despite progress across individual stages of the fuzzing pipeline, existing techniques remain fundamentally constrained when applied to storage systems.
These constraints stem from a persistent mismatch between the assumptions that make fuzzing scalable and the execution realities of storage systems. Table~\ref{tab:fuzzing_limitations} summarizes how these limitations manifest across different stages of the fuzzing pipeline, highlighting that most challenges cut across multiple stages rather than being confined to a single component.

\begin{table}[t]
\centering
\small
\begin{tabular}{lcccc}
\toprule
\textbf{Limitation} & \textbf{Input Gen} & \textbf{Execution} & \textbf{Feedback} & \textbf{Detection} \\
\midrule
Limited Interleaving Control &  & \checkmark & \checkmark &  \\
Deep State \& Aging Limitations     & \checkmark & \checkmark &  &  \\
Semantic Gap in Generation \& Validation & \checkmark &  & \checkmark & \checkmark \\
High Manual Overhead                & \checkmark & \checkmark & \checkmark & \checkmark \\
\bottomrule
\end{tabular}
\caption{Mapping storage-system fuzzing limitations to stages of the fuzzing pipeline.}
\label{tab:fuzzing_limitations}
\end{table}

\subsubsection{Ineffective Control and Exploration of Concurrent Interleavings}
Storage-system testing inherently involves subtle interleavings among concurrent threads or nodes and externally triggered events such as crashes or network perturbations. Existing fuzzers rely largely on the operating system’s default scheduling behavior, offering limited control over thread preemption or execution order within a node, and virtually no control over event timing across nodes. As a result, concurrency-dependent behaviors are explored opportunistically rather than systematically.

Overall, limited controllability of execution interleavings constrains both coverage and reproducibility, and this limitation spans execution and feedback stages of the fuzzing pipeline rather than a single component.

\subsubsection{Limitations in Exploring Storage-System State Space}
Storage systems exhibit rapid, phase-sensitive state changes driven by interactions between foreground operations and asynchronous background activities. Small timing differences can steer execution into distinct internal states, even under similar operation sequences.
In practice, fuzzers lack mechanisms to identify, stabilize, or revisit such intermediate states, making phase-dependent behaviors difficult to explore systematically.

Moreover, many storage bugs arise after deep, history-dependent state accumulation shaped by prolonged execution, repeated maintenance activities, or specific persistent layouts. Treating initial system state as part of the input only partially alleviates this problem, as generating and evolving complex persistent states at scale remains challenging.

This limitation reflects a cross-stage issue affecting input generation and execution control, rather than an isolated weakness of any single fuzzing component.

\subsubsection{Semantic Gaps in Test Generation and Bug Detection}
Effective testing of storage systems requires exercising semantics jointly determined by API invocation sequences and concrete parameter values. In practice, however, most fuzzers approximate semantic validity primarily through producer-consumer and resource-dependency models, which focus on generating executable call sequences rather than modeling how parameter values influence system behavior. As a result, fuzzing exploration remains largely syntactic and lacks guidance toward correctness-critical storage behaviors.

Even when fuzzing reaches complex execution scenarios, automatically determining semantic correctness remains difficult. Storage-system bugs often manifest as delayed or state-dependent violations of durability, ordering, or consistency, rather than as immediate crashes. In the absence of precise and automatically derivable specifications, fuzzers rely on heuristic or oracle-based checks, which are inherently brittle and sensitive to environmental variability.

These semantic gaps jointly limit effective test generation, feedback-guided exploration, and reliable bug detection, cutting across multiple stages of the fuzzing pipeline.

\subsubsection{High Manual Overhead in Practice}
Taken together, these limitations shift a substantial portion of the testing burden from automated mechanisms to human expertise in practice.

At the execution stage, although test execution and environment resets are largely automated, substantial manual effort is still required to make executions reliable and diagnosable. Developers often rely on repeated trial-and-error executions and post hoc analysis to reason about observed failures.

In feedback collection and bug detection, generic signals such as coverage or crashes are insufficient to characterize semantic progress or correctness in storage systems. Practitioners must therefore design system-specific feedback metrics and semantic checkers, and manually interpret ambiguous outcomes—effort that depends heavily on domain expertise and does not scale with automation.

Overall, high manual overhead emerges as a systemic consequence of the preceding limitations, rather than as a shortcoming of individual fuzzing techniques.

\subsection{Summary}
This section examined fuzzing through a storage-centric lens, analyzing how its core pipeline mechanisms interact with the semantic, stateful, and long-horizon challenges of storage systems.
While fuzzing provides powerful automation for input generation and execution, its effectiveness depends on assumptions that frequently misalign with storage-system execution realities.

As a result, fuzzing alone cannot eliminate the fundamental testing challenges identified earlier, but instead exposes where automation breaks down and where additional semantic and state-aware guidance is required.
These observations motivate the discussion in the next section, where we examine how learning-based and AI-augmented techniques may help bridge some of these gaps.

%% file: ai.tex
\section{AI-Driven Opportunities for Storage Fuzzing}\label{sub:ai}
As analyzed in Section~\ref{sub:fuzzing}, existing fuzzing pipelines face fundamental difficulties when applied to storage systems. 
These limitations stem from the difficulty of encoding concurrency, long-horizon state evolution, and semantic correctness intent into scalable, automated fuzzing abstractions. In practice, this results in substantial manual and engineering effort that scales poorly and fails to fundamentally resolve these challenges.

Recent advances in artificial intelligence (AI) offer a complementary set of capabilities that are well aligned with these challenges.
AI excels at extracting structure from complex, noisy, and partially observable systems, particularly when behavior unfolds over time and cannot be easily specified through explicit rules~\cite{sutton2018reinforcement}. These strengths naturally match many of the obstacles encountered in storage-system fuzzing, where execution behavior is shaped by concurrency, background activity, and long-term state evolution. 

This section examines where AI can meaningfully assist storage-system fuzzing—and where it cannot.
Rather than surveying AI-based fuzzers or proposing new architectures, we organize the discussion around recurring challenges identified earlier and analyze how AI-derived abstractions and analyses may provide guidance for fuzzing workflows. To structure this analysis, we use the modeling-to-execution gap as an analytical lens to characterize the disconnect between learning rich models of storage-system behavior and turning those models into executable, controllable fuzzing actions.

\begin{figure*}[t]
	\begin{center}
		\includegraphics[width=1\linewidth]{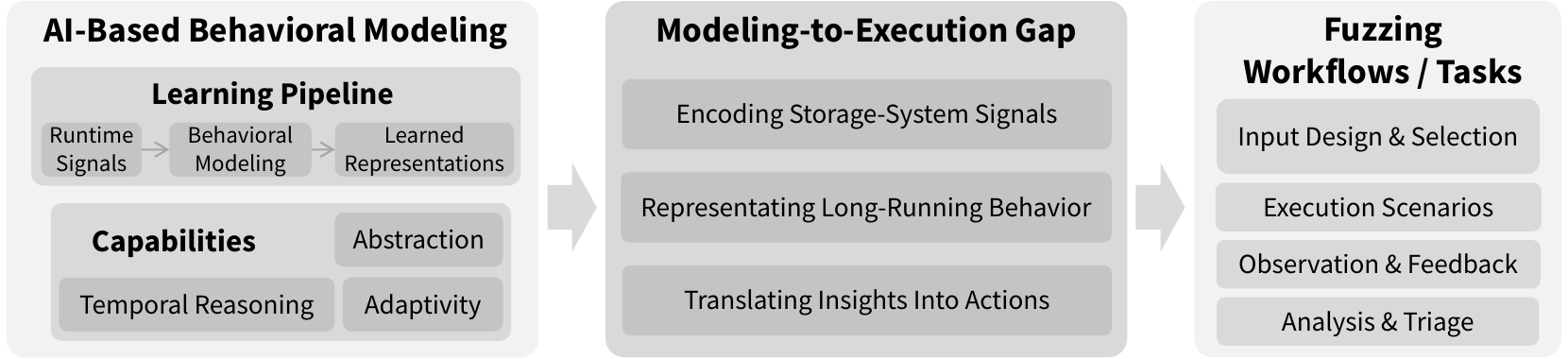} 
	\end{center}
	\caption{AI-based behavioral modeling learns from heterogeneous storage-system signals (left). Bridging the modeling-to-execution gap (center) is required to connect learned models with storage-system fuzzing workflows (right).}\label{fig:ai_learning}
\end{figure*}

\subsection{Capabilities Relevant to Storage Fuzzing}
To understand how AI can complement storage-system fuzzing, we organize the discussion around three roles that learning-based methods can plausibly play: abstraction, temporal reasoning, and adaptive decision support. These roles correspond to how AI can interpret storage-system behavior, reason about its evolution, and inform fuzzing choices.

\textbf{AI for abstraction: interpreting execution signals.  }
Storage systems expose behavior through heterogeneous runtime signals, including logs, traces, performance metrics, and resource usage. In fuzzing, these signals constitute the primary observable evidence of system behavior, yet they are difficult to interpret directly due to their volume, noise, and lack of structure.

Learning-based abstraction enables these heterogeneous signals to be mapped into compact latent representations that summarize execution context and behavioral patterns.
By learning over traces and event sequences, AI models can infer higher-level execution characteristics---such as phases, regimes, or anomalies---from indirect observations~\cite{du2017deeplog, he2021surveyLogAnalysis}. This capability is largely absent in traditional fuzzing, which typically relies on manually defined structures and explicit dependencies.

\textbf{AI for temporal reasoning: modeling long-horizon behavior. }
Many storage-system properties arise from interactions that unfold over extended executions, involving foreground operations, background maintenance, and recovery mechanisms. Such temporal dependencies are difficult to encode explicitly and are often only partially understood by developers.

Temporal learning enables AI models to capture ordering dependencies, phase transitions, and long-term correlations directly from execution histories~\cite{hochreiter1997lstm, sutton2018reinforcement}. Rather than modeling individual events in isolation, these approaches characterize how system behavior evolves over time, providing a view of execution that aligns more closely with long-horizon storage semantics.

\textbf{AI for adaptivity: prioritizing exploration. }
Adaptivity allows learning-based mechanisms to adjust priorities in response to observed behavior, rather than treating all executions uniformly~\cite{sutton2018reinforcement}. In the context of fuzzing, this means that generation strategies, perturbation targets, or checking emphasis can evolve as execution progresses, rather than remaining fixed throughout a testing campaign.
Importantly, AI-driven adaptivity does not replace fuzzing decisions, but can help prioritize where fuzzing effort is spent.
This form of guidance is particularly valuable when the execution space is too large to explore exhaustively, as is common in storage systems.

Together, abstraction, temporal reasoning, and adaptivity illustrate how AI can support storage-system fuzzing across different stages---from interpreting execution signals, to modeling long-horizon behavior, to adapting exploration strategies.
Despite these capabilities, a fundamental challenge remains at the system level: the disconnect between converting heterogeneous storage-system signals into inputs suitable for AI modeling, and translating AI-derived abstractions, temporal models, and adaptive decisions into executable, controllable, and reproducible fuzzing actions. We formalize this challenge next as the modeling-to-execution gap.

\subsection{The Modeling-to-Execution Gap}
Despite these capabilities, applying AI to storage-system fuzzing faces a fundamental challenge: the disconnect between how storage behavior is observed and modeled, and how fuzzing actions are executed.
Storage systems expose behavior through heterogeneous and indirect signals, while learning-based methods operate on abstract representations derived from such signals. Fuzzing, in contrast, requires concrete, controllable, and reproducible execution artifacts. We refer to this disconnect as the modeling-to-execution gap.

As illustrated in the middle of Figure~\ref{fig:ai_learning}, this gap can be viewed along several recurring dimensions that reflect common sources of friction when relating learned models to practical fuzzing workflows:

\begin{itemize}
    \item \textbf{Encoding storage-system signals}: how to transform heterogeneous storage-system signals and attributes into representations suitable for AI modeling.

    \item \textbf{Representing long-running behavior}: how to use learning-based models to identify, characterize, and distinguish long-horizon and multi-phase storage behavior, enabling such states and transitions to be reasoned about at the modeling level.

    \item \textbf{Translating insights into actions}: how to convert abstract model outputs into concrete fuzzing decisions, such as test inputs, execution schedules, or targeted perturbations. 
\end{itemize}
This gap is not a flaw of AI or fuzzing individually, but a consequence of their differing abstractions.
As a result, AI is best viewed as a source of guidance for fuzzing workflows rather than a direct replacement for execution control or correctness validation.

\subsection{Opportunities Along Key Fuzzing Challenges}
Using the modeling-to-execution gap as an analytical lens, we examine how AI techniques may complement storage-system fuzzing across four recurring challenges: concurrency exploration, evolving system state, semantic correctness, and manual modeling effort. Rather than proposing new fuzzing systems or architectures, this discussion highlights where AI-derived representations and analyses can offer additional perspective and guidance for existing fuzzing workflows.

\subsubsection{AI-Guided Exploration of Concurrent Interleavings}
Concurrency interactions remain among the hardest behaviors for fuzzers to explore systematically. Randomized scheduling offers limited control over the space of possible interleavings and execution orderings, often resulting in uneven exploration and poor coverage of concurrency- and phase-sensitive behaviors.

AI-based analysis offers an opportunity to complement fuzzing by exposing structural and temporal characteristics of concurrent executions inferred from execution histories. Rather than controlling scheduling directly, learning-based models can surface recurring interaction patterns among threads, processes, and external events, such as periods of heightened contention, unstable ordering, or phase-dependent interaction regimes.

These patterns provide contextual signals about where concurrency-sensitive behavior tends to concentrate. Such signals can inform fuzzing by highlighting execution regions or interaction patterns that are more likely to exhibit ordering-dependent behavior. Prior work on the analysis of concurrent executions suggests that these temporal and structural patterns are indeed observable from execution traces~\cite{actor, ezaz2026taaf, tehrani2019deeprace}, providing evidence that execution histories contain exploitable structure.

Overall, AI contributes guidance rather than control in the exploration of concurrent behaviors, by revealing structural and temporal context that can help focus attention on concurrency-sensitive execution scenarios that are difficult to reach through unguided scheduling alone.

\subsubsection{AI-Based Understanding of Complex and Evolving System States} ~\label{sub:ai_state}
Storage systems maintain rich internal state that evolves across background activities, maintenance operations, and crash–recovery cycles. In storage systems, many correctness violations depend on internal conditions that are both transient in time and accumulated over long execution histories, making such behaviors difficult for fuzzing to reach and recognize systematically. AI-based analysis offers an opportunity to complement fuzzing by providing additional perspective on how internal state evolves over time.

At a finer granularity, learning-based models can infer latent execution context from externally observable behavior. By learning from black-box signals---such as performance metrics, system-call patterns, or resource-usage traces---AI models can capture abstractions of execution states and transitions that are otherwise opaque to external testing~\cite{du2017deeplog}. These abstractions can help characterize execution phases or state transitions that are more likely to influence downstream behavior or correctness.

At a longer time scale, learning-based models can summarize state evolution across extended executions~\cite{TrailBlazerOSDI20, Jeppu2020ConciseModels}. By capturing progression through the state space over time, AI-derived representations can distinguish shallow executions from those shaped by prolonged background activity, repeated maintenance cycles, or specific persistent layouts, and can highlight classes of accumulated states that remain underexplored.

Taken together, these observations suggest that learning-based representations can serve as a form of state summarization for long-running storage behavior, offering contextual information that may be useful for understanding execution depth and state diversity in fuzzing campaigns. Finally, Appendix A provides an illustrative example of this observation.

\subsubsection{AI-Enhanced Semantic Feedback and Failure Analysis}
Many challenges in testing storage-system correctness depend on exercising semantically relevant inputs---such as API sequences, parameter combinations, and execution context---and on recognizing semantic deviations in system behavior.
However, traditional fuzzing largely operates over syntactic input models and relies on coverage- or fault-based feedback, which provides limited guidance on which inputs are semantically meaningful or how to identify correctness-relevant behaviors beyond crashes.

Beyond exercising syntactically valid executions, learning-based analysis offers an opportunity to better characterize how parameter values and execution context jointly shape semantic behavior. By observing execution outcomes across diverse inputs and real-world usage patterns, learning-based models can capture correlations between parameter choices, execution context, and correctness-sensitive behaviors that are difficult to specify explicitly or capture with handcrafted rules. Such correlations provide insight into semantically meaningful execution scenarios that may be underexplored by syntax- or coverage-driven fuzzing alone.

In addition, learning-based analysis can surface semantic irregularities by modeling relationships across execution artifacts and system layers. By examining correlations among return values, logs, metadata updates, and recovery outcomes, learning-based approaches can highlight executions whose observed behavior appears atypical relative to learned semantic patterns, even in the absence of explicit crashes or faults. These semantic signals complement control-flow coverage by drawing attention to executions that warrant closer inspection from a correctness perspective.

Taken together, these observations suggest that AI-derived semantic signals can complement fuzzing at the input-generation feedback, and bug detection stages, by providing additional context for exploring and analyzing storage-system semantics, rather than serving as explicit correctness oracles. Appendix B presents a lightweight illustrative example.

\subsubsection{AI-Assisted Cross-Layer Modeling and Diagnosis}
A major opportunity for AI-assisted storage fuzzing lies in reducing the substantial manual effort required to model interfaces, maintain specifications, and diagnose failures in complex and evolving storage systems. Much of today’s storage testing practice relies on handcrafted schemas, expert knowledge of cross-layer behavior, and manual interpretation of failure symptoms, all of which are difficult to scale as systems evolve.

At the modeling stage, learning-based techniques enable the automatic inference of interface schemas, argument constraints, and dependency relationships from source code, documentation, and observed execution behavior. By extracting such knowledge automatically, AI can help generate or refine fuzzing schemas without continual manual updates, allowing fuzzing infrastructure to better keep pace with system evolution.

At the diagnostic stage, AI can further reduce the manual effort required after semantic anomalies are detected. Instead of relying solely on developers to manually correlate logs, metadata state, and replay traces across system layers, embedding-based analysis can assist by clustering executions and failures by semantic similarity. Such clustering helps expose recurring failure patterns and reduces the effort needed to triage and interpret complex, cross-layer behaviors. 

Taken together, these modeling and diagnostic capabilities provide guidance for fuzzing workflows by helping prioritize and focus fuzzing effort, highlighting execution regions---such as replay logic, metadata updates, or recovery paths---where targeted perturbations are more likely to expose correctness issues.

\subsubsection{Summary}
AI does not eliminate the fundamental challenges of storage-system fuzzing, but it offers mechanisms to make those challenges more tractable.
By providing abstraction, temporal context, and adaptive guidance, learning-based techniques can complement fuzzing workflows in areas where manual reasoning currently dominates. Framed through the modeling-to-execution gap, these opportunities clarify both the promise and the limits of AI in storage-system fuzzing, highlighting where learning-based insights can guide exploration without replacing execution, control, or validation itself.

%% file: conclusion.tex
\section{Conclusion} \label{sub:conclusion}
This survey examined storage-system testing from a semantic and execution-centric perspective, motivated by the observation that many storage-system failures remain difficult to expose despite decades of testing research. Rather than attributing these gaps to immature tools, we argued that they arise from intrinsic properties of storage-system execution itself, including nondeterministic interleavings, long-horizon state evolution, and correctness semantics that span layers and execution phases.

By organizing existing testing techniques around the failure dimensions and execution properties they target, this survey clarified both their strengths and their fundamental limits. While techniques such as concurrency testing, long-running workloads, crash-consistency analysis, device-level semantic validation, and distributed fault injection each address important classes of failures, no single approach can capture storage-system correctness in isolation. Many violations remain difficult to trigger, observe, or attribute precisely because their causes and effects are separated across time, state, and system layers.

Within this context, we analyzed fuzzing as a promising but fundamentally constrained approach for storage-system testing, highlighting systematic mismatches between conventional fuzzing abstractions and storage-system execution realities. We further discussed how recent advances in AI may help alleviate some of these mismatches by providing abstraction, temporal context, and adaptive guidance, while emphasizing that semantic correctness remains inherently difficult to specify and validate automatically.

Overall, this survey provides a unified framework for understanding why storage systems remain challenging to test in practice and why progress in this space depends not only on more automation, but on better alignment between testing techniques and storage-system semantics. We hope this perspective helps inform future efforts toward more systematic, semantic-aware, and state-aware validation of storage-system correctness.

%% file: appendix.tex
\section{Appendix}

\subsection{A: Illustrative Example of Inferring Storage-System State from External Signals}

This appendix provides a lightweight illustrative example to demonstrate the practical feasibility of inferring internal storage-system states from externally observable signals. The purpose of this example is not to present a complete system, methodology, or evaluation, but to ground the discussion in Section~\ref{sub:ai_state} with concrete observations.

We performed a small exploratory study on LevelDB~\cite{Dean11LevelDB}, focusing on compaction, one of the most resource-intensive and semantically meaningful background activities in LSM-tree systems. Compaction involves sustained read/write bursts, sorting, and file cleanup, and is therefore expected to manifest through distinctive patterns in physical resource usage~\cite{dayan2017monkey}.

\begin{figure}[H]
	\begin{center}
		\includegraphics[width=0.99\linewidth]{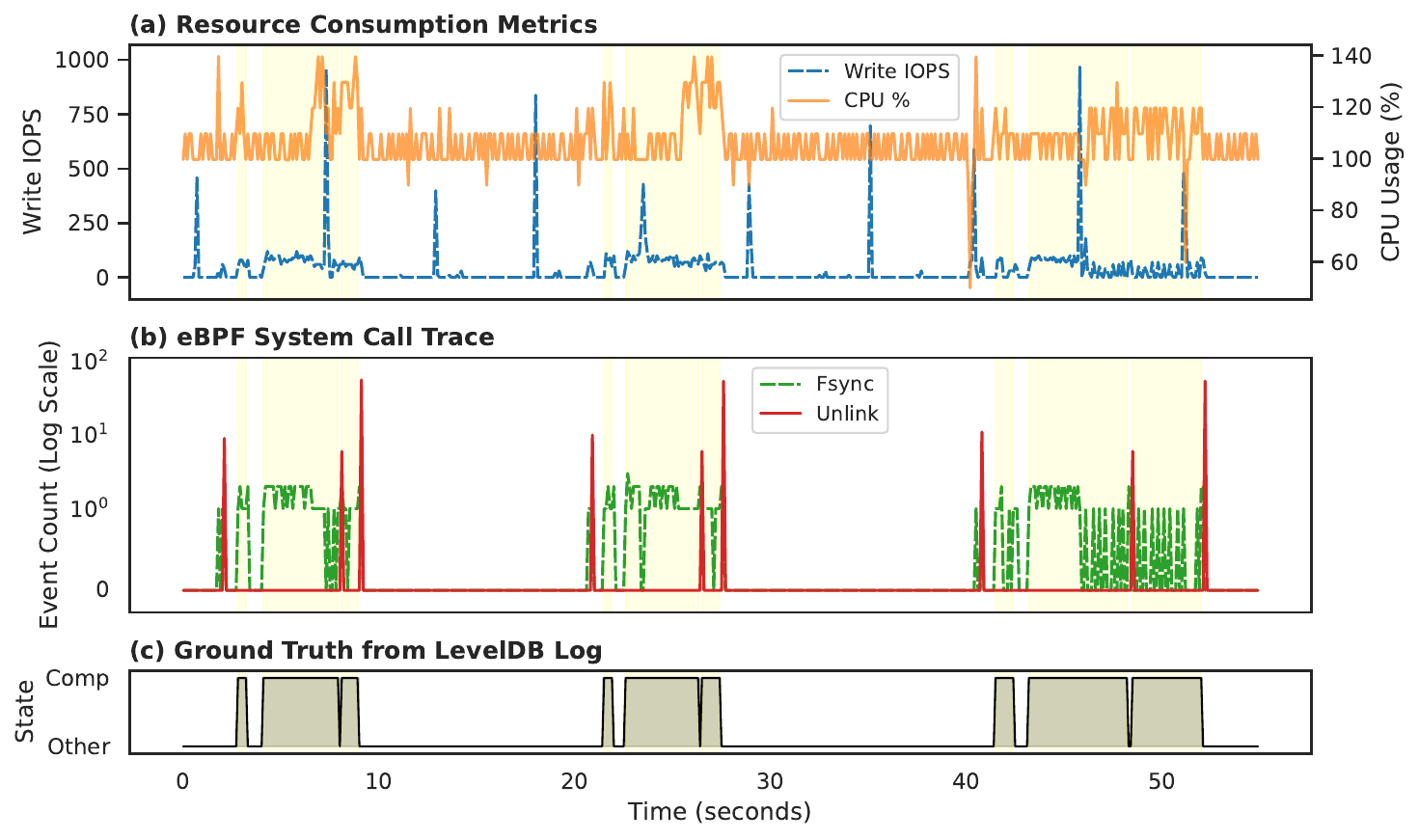} 
	\end{center}
	\caption{\textbf{Temporal correlation between system metrics and LevelDB internal states.} 
    The figure aligns external black-box signals with ground truth labels over a 60-second window. 
    \textbf{(a)} Resource consumption showing sustained Write IOPS and CPU usage during compaction. 
    \textbf{(b)} eBPF traces revealing high-frequency \texttt{fsync} calls and distinct \texttt{unlink} spikes that mark the end of compaction phases. 
    \textbf{(c)} Ground truth compaction intervals derived from internal logs.
    The synchronization between signal bursts (a, b) and state activation (c) confirms that internal states are observable via external metrics.}
    \label{fig:compaction_signature}
\end{figure}

As shown in Figure~\ref{fig:compaction_signature}, these external signals align closely with ground-truth compaction intervals derived from internal logs: sustained write-IOPS and CPU bursts coincide with active merge phases, while clusters of unlink calls sharply delimit phase completion. This alignment illustrates that internal state transitions can be reflected in externally observable metrics, even without internal instrumentation~\cite{Lu14Study}.

Even simple learning-based analyses of such signals can distinguish compaction phases in this setting, indicating that internal storage states “leak” through resource-consumption patterns in a form amenable to automated modeling~\cite{Lu14Study}. While intentionally lightweight and not intended as a performance study, this example supports a broader point of this survey: meaningful internal-state representations can be derived from black-box observations, providing a practical foundation for AI-assisted reasoning about long-lived and evolving system behavior in storage fuzzing.

\subsection{B: Illustrative Prototype for Semantic-Aware Failure Detection
}
This appendix presents a lightweight, illustrative prototype intended to demonstrate the feasibility of learning-based semantic failure detection in storage systems. The goal is not to propose a complete fuzzing framework or provide a comprehensive evaluation, but to show that subtle semantic violations---such as silent durability failures---can be detected from execution artifacts using learning-based analysis.

To investigate the capability of LLMs in discerning semantic violations, we selected LevelDB as our testbed. LevelDB relies on a Write-Ahead Log (WAL) to ensure durability: operations are appended to a log file before being applied to the in-memory MemTable. Crucially, the system supports both asynchronous writes (buffered in the OS page cache) and synchronous writes (using fsync to flush the buffer). Correct recovery relies on prefix consistency: if a synchronous write succeeds, it acts as a durability barrier, guaranteeing that it and all preceding asynchronous writes are persisted~\cite{pillai2015crashconsistency, Rebello2020FsyncFailures} 

To simulate hard-to-detect "silent" durability failures, we instrumented the LevelDB source code to inject stochastic faults. Controlled by a probability parameter, this injection logic intercepts write requests and returns Status::OK to the client while deliberately skipping the underlying WAL write mechanism. This setup emulates real-world scenarios---such as file-system or storage-stack bugs---where the application assumes durability while persistence guarantees are silently violated~\cite{hydra, Rebello2020FsyncFailures}, leaving the system in a corrupted state that traditional crash-based fuzzers may fail to detect.

\begin{figure}[H]
    \centering
    \begin{subfigure}[b]{0.49\textwidth} 
        \centering
        \includegraphics[width=\textwidth]{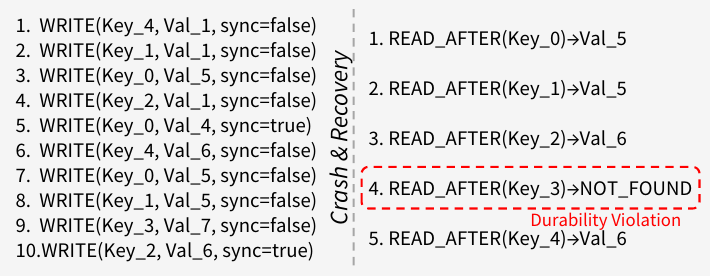}
        \caption{A sample execution trace with a labeled durability violation during the recovery phase.}
        \label{fig:sub-a}
    \end{subfigure}
    \hfill 
    \begin{subfigure}[b]{0.47\textwidth}
        \centering
        \includegraphics[width=\textwidth]{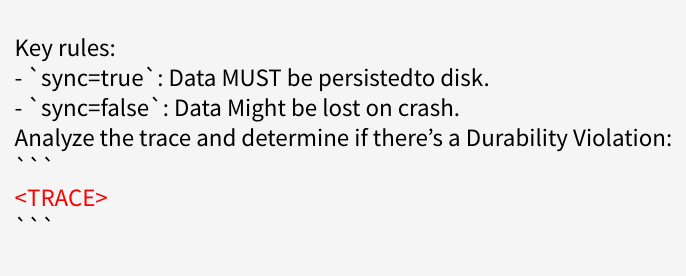}
        \caption{The reasoning rules and prompt template provided to the LLM.}
        \label{fig:sub-b}
    \end{subfigure}
    
    \caption{An example of the execution trace and prompt used for durability verification of LevelDB.}
    \label{fig:prompt_trace}
\end{figure}

We constructed a specialized dataset of execution traces consisting of mixed synchronous and asynchronous Put operations, followed by a simulated crash and a Get phase to inspect the recovered state. We use qwen3-max as the LLM to verify whether the post-crash state respects the linear history of the log~(Figure~\ref{fig:prompt_trace}). We emphasize that this example does not rely on model-specific capabilities; the choice of LLM is representative rather than essential.

The results of this exploratory experiment were encouraging. The LLM was able to reliably distinguish between benign data loss (valid loss of asynchronous tail writes) and critical durability violations (loss of committed data or log-prefix inconsistencies) in this exploratory dataset, highlighting a key advantage over traditional heuristic-based or coverage-guided fuzzers. While conventional tools typically rely on explicit crashes or assertions to flag failures, the LLM effectively acted as a semantic auditor~\cite{Wei2022ChainOfThought}, reasoning over execution histories and recovery outcomes to identify violations that would otherwise remain latent.

Storage semantics also depend on subtle temporal and causal constraints: metadata must reflect durable writes, journal replays must obey ordering, and replicas must maintain invariants over time. These relationships are difficult to encode manually as explicit oracles~\cite{Wu2025Fawkes}, but learning-based models can capture correlations across execution traces, recovery behavior, and observed outcomes, surfacing executions that diverge from expected semantic relationships.

Taken together, this illustrative example demonstrates the feasibility of semantic-aware failure detection using learning-based analysis. While preliminary and intentionally scoped, it reinforces the central argument of this survey: AI-assisted techniques can provide semantic feedback that traditional fuzzers lack, directly addressing latent durability, recovery, and cross-layer correctness challenges discussed in Section~\ref{sub:fuzzing}.




%% file: references.bib
@inproceedings{mock,
  author       = {Jiacheng Xu and
                  Xuhong Zhang and
                  Shouling Ji and
                  Yuan Tian and
                  Binbin Zhao and
                  Qinying Wang and
                  Peng Cheng and
                  Jiming Chen},
  title        = {{MOCK:} Optimizing Kernel Fuzzing Mutation with Context-aware Dependency},
  booktitle    = {31st Annual Network and Distributed System Security Symposium, {NDSS}
                  2024, San Diego, California, USA, February 26 - March 1, 2024},
  publisher    = {The Internet Society},
  year         = {2024},
  url          = {https://www.ndss-symposium.org/ndss-paper/mock-optimizing-kernel-fuzzing-mutation-with-context-aware-dependency/},
  bibsource    = {dblp computer science bibliography, https://dblp.org}
}

@inproceedings{hydra,
  author       = {Seulbae Kim and
                  Meng Xu and
                  Sanidhya Kashyap and
                  Jungyeon Yoon and
                  Wen Xu and
                  Taesoo Kim},
  editor       = {Tim Brecht and
                  Carey Williamson},
  title        = {Finding semantic bugs in file systems with an extensible fuzzing framework},
  booktitle    = {Proceedings of the 27th {ACM} Symposium on Operating Systems Principles,
                  {SOSP} 2019, Huntsville, ON, Canada, October 27-30, 2019},
  pages        = {147--161},
  publisher    = {{ACM}},
  year         = {2019},
  url          = {https://doi.org/10.1145/3341301.3359662},
  doi          = {10.1145/3341301.3359662},
  bibsource    = {dblp computer science bibliography, https://dblp.org}
}

@inproceedings{janus,
  author       = {Wen Xu and
                  Hyungon Moon and
                  Sanidhya Kashyap and
                  Po{-}Ning Tseng and
                  Taesoo Kim},
  title        = {Fuzzing File Systems via Two-Dimensional Input Space Exploration},
  booktitle    = {2019 {IEEE} Symposium on Security and Privacy, {SP} 2019, San Francisco,
                  CA, USA, May 19-23, 2019},
  pages        = {818--834},
  publisher    = {{IEEE}},
  year         = {2019},
  url          = {https://doi.org/10.1109/SP.2019.00035},
  doi          = {10.1109/SP.2019.00035},
  bibsource    = {dblp computer science bibliography, https://dblp.org}
}

@inproceedings{healer,
  author       = {Hao Sun and
                  Yuheng Shen and
                  Cong Wang and
                  Jianzhong Liu and
                  Yu Jiang and
                  Ting Chen and
                  Aiguo Cui},
  editor       = {Robbert van Renesse and
                  Nickolai Zeldovich},
  title        = {{HEALER:} Relation Learning Guided Kernel Fuzzing},
  booktitle    = {{SOSP} '21: {ACM} {SIGOPS} 28th Symposium on Operating Systems Principles,
                  Virtual Event / Koblenz, Germany, October 26-29, 2021},
  pages        = {344--358},
  publisher    = {{ACM}},
  year         = {2021},
  url          = {https://doi.org/10.1145/3477132.3483547},
  doi          = {10.1145/3477132.3483547},
  bibsource    = {dblp computer science bibliography, https://dblp.org}
}

@inproceedings{monarch,
  author       = {Tao Lyu and
                  Liyi Zhang and
                  Zhiyao Feng and
                  Yueyang Pan and
                  Yujie Ren and
                  Meng Xu and
                  Mathias Payer and
                  Sanidhya Kashyap},
  editor       = {Saurabh Bagchi and
                  Yiying Zhang},
  title        = {Monarch: {A} Fuzzing Framework for Distributed File Systems},
  booktitle    = {Proceedings of the 2024 {USENIX} Annual Technical Conference, {USENIX}
                  {ATC} 2024, Santa Clara, CA, USA, July 10-12, 2024},
  pages        = {529--543},
  publisher    = {{USENIX} Association},
  year         = {2024},
  url          = {https://www.usenix.org/conference/atc24/presentation/lyu},
  bibsource    = {dblp computer science bibliography, https://dblp.org}
}

@inproceedings{actor,
  author       = {Marius Fleischer and
                  Dipanjan Das and
                  Priyanka Bose and
                  Weiheng Bai and
                  Kangjie Lu and
                  Mathias Payer and
                  Christopher Kruegel and
                  Giovanni Vigna},
  editor       = {Joseph A. Calandrino and
                  Carmela Troncoso},
  title        = {{ACTOR:} Action-Guided Kernel Fuzzing},
  booktitle    = {32nd {USENIX} Security Symposium, {USENIX} Security 2023, Anaheim,
                  CA, USA, August 9-11, 2023},
  pages        = {5003--5020},
  publisher    = {{USENIX} Association},
  year         = {2023},
  url          = {https://www.usenix.org/conference/usenixsecurity23/presentation/fleischer},
  bibsource    = {dblp computer science bibliography, https://dblp.org}
}

@inproceedings{krace,
  author       = {Meng Xu and
                  Sanidhya Kashyap and
                  Hanqing Zhao and
                  Taesoo Kim},
  title        = {Krace: Data Race Fuzzing for Kernel File Systems},
  booktitle    = {2020 {IEEE} Symposium on Security and Privacy, {SP} 2020, San Francisco,
                  CA, USA, May 18-21, 2020},
  pages        = {1643--1660},
  publisher    = {{IEEE}},
  year         = {2020},
  url          = {https://doi.org/10.1109/SP40000.2020.00078},
  doi          = {10.1109/SP40000.2020.00078},
  bibsource    = {dblp computer science bibliography, https://dblp.org}
}

@inproceedings{horcrux,
  author       = {Fuchen Ma and
                  Yuanliang Chen and
                  Yuanhang Zhou and
                  Zhen Yan and
                  Hao Sun and
                  Yu Jiang},
  editor       = {Lujo Bauer and
                  Giancarlo Pellegrino},
  title        = {Finding Metadata Inconsistencies in Distributed File Systems via Cross-Node
                  Operation Modeling},
  booktitle    = {34th {USENIX} Security Symposium, {USENIX} Security 2025, Seattle,
                  WA, USA, August 13-15, 2025},
  pages        = {7525--7543},
  publisher    = {{USENIX} Association},
  year         = {2025},
  url          = {https://www.usenix.org/conference/usenixsecurity25/presentation/ma-fuchen},
  bibsource    = {dblp computer science bibliography, https://dblp.org}
}

@article{libfuzzer,
  author       = {Valentin J. M. Man{\`{e}}s and
                  HyungSeok Han and
                  Choongwoo Han and
                  Sang Kil Cha and
                  Manuel Egele and
                  Edward J. Schwartz and
                  Maverick Woo},
  title        = {The Art, Science, and Engineering of Fuzzing: {A} Survey},
  journal      = {{IEEE} Trans. Software Eng.},
  volume       = {47},
  number       = {11},
  pages        = {2312--2331},
  year         = {2021},
  url          = {https://doi.org/10.1109/TSE.2019.2946563},
  doi          = {10.1109/TSE.2019.2946563},
  bibsource    = {dblp computer science bibliography, https://dblp.org}
}

@inproceedings{grayc,
  author       = {Karine Even{-}Mendoza and
                  Arindam Sharma and
                  Alastair F. Donaldson and
                  Cristian Cadar},
  editor       = {Ren{\'{e}} Just and
                  Gordon Fraser},
  title        = {GrayC: Greybox Fuzzing of Compilers and Analysers for {C}},
  booktitle    = {Proceedings of the 32nd {ACM} {SIGSOFT} International Symposium on
                  Software Testing and Analysis, {ISSTA} 2023, Seattle, WA, USA, July
                  17-21, 2023},
  pages        = {1219--1231},
  publisher    = {{ACM}},
  year         = {2023},
  url          = {https://doi.org/10.1145/3597926.3598130},
  doi          = {10.1145/3597926.3598130},
  bibsource    = {dblp computer science bibliography, https://dblp.org}
}

@inproceedings{restler,
  author       = {Vaggelis Atlidakis and
                  Patrice Godefroid and
                  Marina Polishchuk},
  editor       = {Joanne M. Atlee and
                  Tevfik Bultan and
                  Jon Whittle},
  title        = {RESTler: stateful {REST} {API} fuzzing},
  booktitle    = {Proceedings of the 41st International Conference on Software Engineering,
                  {ICSE} 2019, Montreal, QC, Canada, May 25-31, 2019},
  pages        = {748--758},
  publisher    = {{IEEE} / {ACM}},
  year         = {2019},
  url          = {https://doi.org/10.1109/ICSE.2019.00083},
  doi          = {10.1109/ICSE.2019.00083},
  bibsource    = {dblp computer science bibliography, https://dblp.org}
}

@inproceedings{nautilus,
  author       = {Cornelius Aschermann and
                  Tommaso Frassetto and
                  Thorsten Holz and
                  Patrick Jauernig and
                  Ahmad{-}Reza Sadeghi and
                  Daniel Teuchert},
  title        = {{NAUTILUS:} Fishing for Deep Bugs with Grammars},
  booktitle    = {26th Annual Network and Distributed System Security Symposium, {NDSS}
                  2019, San Diego, California, USA, February 24-27, 2019},
  publisher    = {The Internet Society},
  year         = {2019},
  url          = {https://www.ndss-symposium.org/ndss-paper/nautilus-fishing-for-deep-bugs-with-grammars/},
  bibsource    = {dblp computer science bibliography, https://dblp.org}
}

@misc{syzkaller,
  author       = {Google},
  title        = {syzkaller - kernel fuzzer},
  howpublished = {\url{https://github.com/google/syzkaller}},
  year         = {2025},
  note         = {Accessed: 2025-12-23}
}

@misc{afl,
  author       = {Michal Zalewski},
  title        = {american fuzzy lop},
  howpublished = {\url{https://lcamtuf.coredump.cx/afl/}},
  year         = {2017},
  note         = {Accessed: 2025-12-23}
}

@misc{grammar-mutator,
  author       = {Shengtuo Hu},
  title        = {Grammar Mutator - AFL++},
  howpublished = {\url{https://github.com/AFLplusplus/Grammar-Mutator}},
  year         = {2024},
  note         = {Accessed: 2025-12-23}
}

@techreport{fsltraces,
  title        = {File Systems and Storage Lab (FSL) I/O Trace Repository},
  author       = {Zadok, Erez and et al.},
  institution  = {Stony Brook University, File Systems and Storage Lab},
  year         = {2005},
  note         = {Historical I/O traces previously published by FSL. Repository no longer publicly hosted.},
  url          = {https://www.fsl.cs.stonybrook.edu/}
}

@inproceedings{msrtraces,
  title        = {Write Off-Loading: Practical Power Management for Enterprise Storage},
  author       = {Narayanan, Dushyanth and Donnelly, Austin and Rowstron, Antony},
  booktitle    = {Proceedings of the 6th USENIX Conference on File and Storage Technologies (FAST'08)},
  pages        = {253--267},
  year         = {2008},
  organization = {USENIX Association}
}

@inproceedings{ycsb,
  title={Benchmarking cloud serving systems with YCSB},
  author={Cooper, Brian F and Silberstein, Adam and Tam, Erwin and Ramakrishnan, Raghu and Sears, Russell},
  booktitle={Proceedings of the 1st ACM Symposium on Cloud Computing (SoCC)},
  pages={143--154},
  year={2010},
  organization={ACM}
}

@techreport{tpcc,
  title        = {{TPC-C} Benchmark},
  institution  = {Transaction Processing Performance Council (TPC)},
  year         = {2010},
  howpublished = {\url{http://www.tpc.org/tpcc/}},
  note         = {Accessed: 2025-02-XX}
}

@misc{rocksdb_stress,
  title        = {db\_stress: RocksDB Stress and Correctness Testing Tool},
  author       = {Facebook Engineering},
  howpublished = {\url{https://github.com/facebook/rocksdb/blob/main/db_stress_tool/db_stress.cc}},
  note         = {Part of the RocksDB test suite. Accessed: 2025-12-03},
  year         = {2025}
}

@article{crashmonkey,
  title        = {CrashMonkey and ACE: Systematically Testing File-System Crash Consistency},
  author       = {Mohan, Jayashree and Martinez, Ashlie and Ponnapalli, Soujanya and Raju, Pandian and Chidambaram, Vijay},
  journal      = {ACM Transactions on Storage (TOS)},
  volume       = {15},
  number       = {2},
  pages        = {14:1--14:34},
  year         = {2019},
  doi          = {10.1145/3320275},
  url          = {https://www.microsoft.com/en-us/research/wp-content/uploads/2021/10/tos-crashmonkey.pdf}
}

@misc{xfstests,
  title        = {{xfstests}: File System Test Suite},
  howpublished = {\url{https://git.kernel.org/pub/scm/fs/xfs/xfstests-dev.git/}},
  note         = {Includes the \texttt{fsstress} tool. Accessed: 2025-02-XX},
  year         = {2025},
  organization = {Kernel.org}
}

@techreport{landslide,
  title={{Landslide}: Systematic Concurrency Testing for {PThreads} and the {Linux} Kernel},
  author={Lal, Akash and Lu, Shan and Cadar, Cristian and Reps, Thomas and Engler, Dawson and Fonseca, Rodrigo},
  institution={Carnegie Mellon University},
  number={CMU-CS-14-105},
  year={2014},
  url={https://reports-archive.adm.cs.cmu.edu/anon/2014/CMU-CS-14-105.pdf}
}

@inproceedings{mohan2018b3,
  title        = {Finding Crash-Consistency Bugs with Bounded Black-Box Crash Testing},
  author       = {Mohan, Jayashree and Martinez, Ashlie and Ponnapalli, Soujanya and Raju, Pandian and Chidambaram, Vijay},
  booktitle    = {13th USENIX Symposium on Operating Systems Design and Implementation (OSDI 2018)},
  pages        = {33--50},
  year         = {2018},
  url          = {https://www.cs.utexas.edu/~vijay/papers/osdi18-crashmonkey.pdf}
}

@misc{fsverity-doc,
  author       = {{Linux Kernel Project}},
  title        = {fs-verity: Read-Only File-Based Authenticity Protection},
  howpublished = {\url{https://docs.kernel.org/filesystems/fsverity.html}},
  note         = {Accessed December 2025}
}

@misc{fsck-manpage,
  author       = {{Linux man-pages Project}},
  title        = {fsck(8): Check and Repair a Linux Filesystem},
  howpublished = {\url{https://man7.org/linux/man-pages/man8/fsck.8.html}},
  note         = {Accessed December 2025}
}

@misc{jepsen,
  author       = {Kingsbury, Kyle},
  title        = {Jepsen: Distributed Systems Safety Verification},
  howpublished = {\url{https://jepsen.io/}},
  note         = {Accessed: 2025-12-03}
}

@inproceedings{perennial,
  title        = {Verifying concurrent, crash-safe systems with Perennial},
  author       = {Chajed, Tej and Tassarotti, Joseph and Kaashoek, M. Frans and Zeldovich, Nickolai},
  booktitle    = {Proceedings of the 27th ACM Symposium on Operating Systems Principles (SOSP)},
  year         = {2019},
  pages        = {698--712},
  publisher    = {ACM},
  doi          = {10.1145/3341301.3359632},
  url          = {https://doi.org/10.1145/3341301.3359632},
}

@inproceedings{rprc2025,
  title     = {Runtime Protocol Refinement Checking for Distributed Protocol Implementations},
  author    = {Ding, Yilun and Liu, Xiyue and Zhang, Sihan and Hsiao, Hengrui and Joshi, Pallavi and Chai, Wei},
  booktitle = {Proceedings of the 22nd USENIX Symposium on Networked Systems Design and Implementation (NSDI 2025)},
  year      = {2025},
  publisher = {USENIX Association},
  note      = {To appear},
}

@inproceedings{flashsim,
  author    = {Li, Jie and Grupp, Lloyd P. and Swanson, Steven and Siegel, Alan},
  title     = {FlashSim: A Simulator for NAND Flash-based Solid-State Drives},
  booktitle = {Proceedings of the 2009 IEEE International Symposium on Modeling, Analysis and Simulation of Computer and Telecommunication Systems (MASCOTS)},
  year      = {2009},
  pages     = {1--10},
  publisher = {IEEE},
  doi       = {10.1109/MASCOT.2009.5366154}
}

@misc{pmreorder,
  title        = {pmreorder: The Persistent Memory Store-Reordering Tool},
  author       = {{PMDK / Intel Corporation}},
  howpublished = {\url{https://pmem.io/pmdk/pmreorder/}},
  note         = {Accessed: 2025-12-03}
}

@inproceedings{xfdetector,
  author    = {Liu, Sihang and Seemakhupt, Korakit and Wei, Yizhou and Wenisch, Thomas F. and Kolli, Aasheesh and Khan, Samira},
  title     = {Cross-Failure Bug Detection in Persistent Memory Programs},
  booktitle = {Proceedings of the 25th ACM International Conference on Architectural Support for Programming Languages and Operating Systems (ASPLOS '20)},
  year      = {2020},
  publisher = {ACM}
}

@inproceedings{pmtest,
  author    = {Liu, Sihang and Wei, Yizhou and Zhao, Jishen and Kolli, Aasheesh and Khan, Samira},
  title     = {PMTest: A Fast and Flexible Testing Framework for Persistent Memory Programs},
  booktitle = {Proceedings of the Twenty-Fourth International Conference on Architectural Support for Programming Languages and Operating Systems (ASPLOS)},
  year      = {2019},
  pages     = {411--425},
  publisher = {ACM},
  doi       = {10.1145/3297858.3304015}
}

@inproceedings {mqsim,
author = {Arash Tavakkol and Juan G{\'o}mez-Luna and Mohammad Sadrosadati and Saugata Ghose and Onur Mutlu},
title = {{MQSim}: A Framework for Enabling Realistic Studies of Modern {Multi-Queue} {SSD} Devices},
booktitle = {16th USENIX Conference on File and Storage Technologies (FAST 18)},
year = {2018},
isbn = {978-1-931971-42-3},
address = {Oakland, CA},
pages = {49--66},
url = {https://www.usenix.org/conference/fast18/presentation/tavakkol},
publisher = {USENIX Association},
month = feb
}

@inproceedings{mocket,
  title        = {Model Checking Guided Testing for Distributed Systems},
  author       = {Wang, Dong and Dou, Wensheng and Gao, Yu and Wu, Chenao and Wei, Jun and Huang, Tao},
  booktitle    = {Proceedings of the 18th European Conference on Computer Systems (EuroSys ’23)},
  year         = {2023},
  publisher    = {ACM},
  doi          = {10.1145/3552326.3587442},
  url          = {https://doi.org/10.1145/3552326.3587442},
}

@inproceedings{samc,
  title        = {SAMC: A Lightweight Symbolic Model Checker for Distributed Systems},
  author       = {Chen, Bei and Zhong, Zhiqiang and Wang, Yanlin and Zhai, Jiannan and Zhang, Yu and Chen, Haibo and Li, Jinyang},
  booktitle    = {Proceedings of the 12th USENIX Symposium on Networked Systems Design and Implementation (NSDI)},
  year         = {2015},
  pages        = {399--414},
  publisher    = {USENIX Association},
  address      = {Oakland, CA},
  url          = {https://www.usenix.org/conference/nsdi15/technical-sessions/presentation/chen},
}

@inproceedings{molly,
  author    = {Alvaro, Peter and Conway, Neil and Hellerstein, Joseph M. and Sears, Russell},
  title     = {Lineage-driven Fault Injection},
  booktitle = {Proceedings of the 2015 ACM SIGMOD International Conference on Management of Data},
  year      = {2015},
  pages     = {331--346},
  publisher = {Association for Computing Machinery},
  address   = {Melbourne, Victoria, Australia},
  doi       = {10.1145/2723372.2723711},
  url       = {https://dl.acm.org/doi/10.1145/2723372.2723711}
}

@inproceedings{iaso,
  title        = {IASO: A Fail-Slow Detection Framework for Cloud Storage Services},
  author       = {Panda, Aishwarya and He, Jiayi and Teodorescu, Raluca A. and Kocoloski, Brent and Min, Chul},
  booktitle    = {Proceedings of the 17th USENIX Symposium on Networked Systems Design and Implementation (NSDI)},
  year         = {2020},
  pages        = {361--378},
  publisher    = {USENIX Association},
  address      = {Santa Clara, CA},
  url          = {https://www.usenix.org/conference/nsdi20/presentation/panda},
}

@inproceedings{timeoutstudy2018,
  author    = {Ting Dai and Jingzhu He and Xiaohui Gu and Shan Lu},
  title     = {Understanding Real-World Timeout Problems in Cloud Server Systems},
  booktitle = {Proceedings of the 2018 IEEE International Conference on Cloud Engineering (IC2E)},
  year      = {2018},
  pages     = {1--11},
  publisher = {IEEE Computer Society},
  address   = {Orlando, FL, USA},
  doi       = {10.1109/IC2E.2018.00022},
  url       = {https://doi.org/10.1109/IC2E.2018.00022},
}

@inproceedings{neat,
  title        = {An Analysis of Network-Partitioning Failures in Cloud Systems},
  author       = {Alquraan, Mohammad and Sharma, Sonam and Gupta, Liting Hu and Albarghouthi, Aws and Mutlu, Onur and Gupta, Indranil},
  booktitle    = {Proceedings of the 13th USENIX Symposium on Operating Systems Design and Implementation (OSDI)},
  year         = {2018},
  pages        = {51--68},
  publisher    = {USENIX Association},
  address      = {Carlsbad, CA},
}

@misc{rocksdb-github,
  author       = {{Meta Platforms, Inc.}},
  title        = {RocksDB: A Persistent Key-Value Store for Fast Storage},
  howpublished = {\url{https://github.com/facebook/rocksdb}},
  note         = {Accessed December 2025}
}

@inproceedings{fscq,
  author    = {Zhong, Xi Wang and Kaashoek, M. Frans and Zeldovich, Nickolai},
  title     = {FSCQ: A Crash-Safe File System Verified with Coq},
  booktitle = {Proceedings of the 12th USENIX Symposium on Operating Systems Design and Implementation (OSDI '15)},
  year      = {2015},
  pages     = {483--498},
  publisher = {USENIX Association},
  url       = {https://www.usenix.org/conference/osdi15/technical-sessions/presentation/zhang_xi}
}

@inproceedings{ccfs,
  author    = {Pillai, Padmanabhan and Krishnan, Mahesh and Kolli, Ashish and Zhang, Yandong and Zhang, Kaiyuan and Zhou, Yuanyuan and Yang, Junfeng},
  title     = {Application Crash Consistency and Performance with CCFS},
  booktitle = {Proceedings of the 15th USENIX Conference on File and Storage Technologies (FAST '17)},
  year      = {2017},
  pages     = {203--216},
  publisher = {USENIX Association},
  url       = {https://www.usenix.org/conference/fast17/technical-sessions/presentation/pillai}
}

@inproceedings{zfs-scrub,
  title={ZFS: The Last Word in File Systems},
  author={Leventhal, Adam H.},
  booktitle={Proc. of SNIA Storage Developer Conference},
  year={2007}
}

@misc{dm-verity,
  title        = {dm-verity - The Linux Kernel documentation},
  howpublished = {\url{https://docs.kernel.org/admin-guide/device-mapper/verity.html}},
  note         = {Accessed: 2025-12-03},
  year         = {2025}
}

@inproceedings{chess,
  title={Finding and reproducing Heisenbugs in concurrent programs},
  author={Musuvathi, Madanlal and Qadeer, Shaz},
  booktitle={Proceedings of the 8th USENIX Symposium on Operating Systems Design and Implementation (OSDI)},
  pages={267--280},
  year={2008},
  organization={USENIX Association}
}

@inproceedings{Fang2023HeteroScore,
  author    = {Chongzhou Fang and Najmeh Nazari and Behnam Omidi and Han Wang and Aditya Puri and Manish Arora and Setareh Rafatirad and Houman Homayoun and Khaled N. Khasawneh},
  title     = {HeteroScore: Evaluating and Mitigating Cloud Security Threats Brought by Heterogeneity},
  booktitle = {30th Annual Network and Distributed System Security Symposium (NDSS ’23)},
  year      = {2023},
  doi       = {10.14722/ndss.2023.24996},
  url       = {https://doi.org/10.14722/ndss.2023.24996}
}

@inproceedings{Chen2023PushButtonRainmaker,
  author    = {Yinfang Chen and Xudong Sun and Suman Nath and Ze Yang and Tianyin Xu},
  title     = {Push-Button Reliability Testing for Cloud-Backed Applications with Rainmaker},
  booktitle = {20th USENIX Symposium on Networked Systems Design and Implementation (NSDI ’23)},
  year      = {2023},
  isbn      = {978-1-939133-33-5},
  address   = {Boston, MA, USA},
  pages     = {1701--1716},
  url       = {https://www.usenix.org/conference/nsdi23/presentation/chen-yinfang},
  publisher = {USENIX Association},
}

@article{Lu2022CloudRaid,
  author    = {Jie Lu and Feng Li and Chen Liu and Lian Li and Xiaobing Feng and Jingling Xue},
  title     = {CloudRaid: Detecting Distributed Concurrency Bugs via Log Mining and Enhancement},
  journal   = {IEEE Transactions on Software Engineering},
  volume    = {48},
  number    = {2},
  pages     = {662--677},
  year      = {2022},
  doi       = {10.1109/TSE.2020.2978394},
}

@inproceedings{Zhai2020CheckBeforeYouChange,
  author    = {Ennan Zhai and Ang Chen and Ruzica Piskac and Mahesh Balakrishnan and Bingchuan Tian and Bo Song and Haoliang Zhang},
  title     = {Check before You Change: Preventing Correlated Failures in Service Updates},
  booktitle = {17th USENIX Symposium on Networked Systems Design and Implementation (NSDI ’20)},
  year      = {2020},
  isbn      = {978-1-939133-13-7},
  address   = {Santa Clara, CA, USA},
  pages     = {575–589},
  url       = {https://www.usenix.org/conference/nsdi20/presentation/zhai},
}

@inproceedings{Dai2018TimeoutProblems,
  author    = {Ting Dai and Jingzhu He and Xiaohui Gu and Shan Lu},
  title     = {Understanding Real-World Timeout Problems in Cloud Server Systems},
  booktitle = {2018 IEEE International Conference on Cloud Engineering (IC2E 2018)},
  year      = {2018},
  pages     = {1--11},
  location  = {Orlando, FL, USA},
  publisher = {IEEE Computer Society},
  url       = {https://doi.org/10.1109/IC2E.2018.00022},
  doi       = {10.1109/IC2E.2018.00022},
}

@Article{Rodeh13BTRFS,
  author        = {Rodeh, Ohad and Bacik, Josef and Mason, Chris},
  title         = {BTRFS: The Linux B-Tree Filesystem},
  journal       = {Trans. Storage},
  year          = {2013},
  volume        = {9},
  number        = {3},
  pages         = {9:1--9:32},
  month         = aug,
  note = {https::/doi/org/10.1145/2501620.2501623},
  issn          = {1553-3077},
  __markedentry = {[wang:1]},
  abstract      = {BTRFS is a Linux filesystem that has been adopted as the default filesystem in some popular versions of Linux. It is based on copy-on-write, allowing for efficient snapshots and clones. It uses B-trees as its main on-disk data structure. The design goal is to work well for many use cases and workloads. To this end, much effort has been directed to maintaining even performance as the filesystem ages, rather than trying to support a particular narrow benchmark use-case. Linux filesystems are installed on smartphones as well as enterprise servers. This entails challenges on many different fronts. ---Scalability. The filesystem must scale in many dimensions: disk space, memory, and CPUs. ---Data integrity. Losing data is not an option, and much effort is expended to safeguard the content. This includes checksums, metadata duplication, and RAID support built into the filesystem. ---Disk diversity. The system should work well with SSDs and hard disks. It is also expected to be able to use an array of different sized disks, which poses challenges to the RAID and striping mechanisms. This article describes the core ideas, data structures, and algorithms of this filesystem. It sheds light on the challenges posed by defragmentation in the presence of snapshots, and the tradeoffs required to maintain even performance in the face of a wide spectrum of workloads.},
  acmid         = {2501623},
  address       = {New York, NY, USA},
  articleno     = {9},
  numpages      = {32},
  publisher     = {ACM},

}

@inproceedings{Atlidakis16POSIX,
	__Markedentry = {[wang:1]},
	Acmid = {2901350},
	Address = {New York, NY, USA},
	Articleno = {19},
	Author = {Atlidakis, Vaggelis and Andrus, Jeremy and Geambasu, Roxana and Mitropoulos, Dimitris and Nieh, Jason},
	Booktitle = {Proceedings of the Eleventh European Conference on Computer Systems},
	Doi = {10.1145/2901318.2901350},
	Isbn = {978-1-4503-4240-7},
	Location = {London, United Kingdom},
	Numpages = {17},
	Pages = {19:1--19:17},
	Publisher = {ACM},
	Series = {EuroSys '16},
	Title = {POSIX Abstractions in Modern Operating Systems: The Old, the New, and the Missing},
	Url = {http://doi.acm.org/10.1145/2901318.2901350},
	Year = {2016},
	Bdsk-Url-1 = {http://doi.acm.org/10.1145/2901318.2901350},
	Bdsk-Url-2 = {https://dx.doi.org/10.1145/2901318.2901350}}

@inproceedings{Ghemawat03Google,
	Acmid = {945450},
	Address = {New York, NY, USA},
	Author = {Ghemawat, Sanjay and Gobioff, Howard and Leung, Shun-Tak},
	Booktitle = {Proceedings of the Nineteenth ACM Symposium on Operating Systems Principles},
	Doi = {10.1145/945445.945450},
	Isbn = {1-58113-757-5},
	Location = {Bolton Landing, NY, USA},
	Numpages = {15},
	Pages = {29--43},
	Publisher = {ACM},
	Series = {SOSP '03},
	Title = {The Google File System},
	Url = {http://doi.acm.org/10.1145/945445.945450},
	Year = {2003}}

@article{Lu14Study,
	Acmid = {2560012},
	Address = {New York, NY, USA},
	Articleno = {3},
	Author = {Lu, Lanyue and Arpaci-Dusseau, Andrea C. and Arpaci-Dusseau, Remzi H. and Lu, Shan},
	Doi = {10.1145/2560012},
	Issn = {1553-3077},
	Issue_Date = {January 2014},
	Journal = {Trans. Storage},
	Month = jan,
	Number = {1},
	Numpages = {32},
	Pages = {3:1--3:32},
	Publisher = {ACM},
	Title = {A Study of Linux File System Evolution},
	Url = {http://doi.acm.org/10.1145/2560012},
	Volume = {10},
	Year = {2014}}

@InProceedings{JianXu16NOVA,
  author    = {Jian Xu and Steven Swanson},
  title     = {{NOVA}: A Log-structured File System for Hybrid Volatile/Non-volatile Main Memories},
  booktitle = {14th {USENIX} Conference on File and Storage Technologies},
  year      = {2016},
  series    = {FAST'16},
  pages     = {323--338},
  publisher = {{USENIX} Association},
  isbn      = {978-1-931971-28-7},
  note = {https::/doi/org/10.5555/2930583.2930608},
  url       = {https://www.usenix.org/conference/fast16/technical-sessions/presentation/xu},
}

@misc{FileBench,
  author       = {{Filebench Project}},
  title        = {Filebench: A File System Workload Generator},
  howpublished = {\url{https://github.com/filebench/filebench}},
  note         = {Accessed January 2025}
}

@misc{fio,
  author       = {Axboe, Jens},
  title        = {fio: Flexible I/O Tester},
  howpublished = {\url{https://github.com/axboe/fio}},
  note         = {Accessed January 2025}
}

@inproceedings{ext4,
  title={The ext4 filesystem},
  author={Mathur, Avantika and Cao, Mingming and Bhattacharya, Suparna and Dilger, Theodore Tso and Tomas, Andreas},
  booktitle={Proceedings of the Linux Symposium},
  volume={2},
  pages={21--33},
  year={2007}
}

@inproceedings{Lee15F2FS,
	Acmid = {2750503},
	Author = {Lee, Changman and Sim, Dongho and Hwang, Joo-Young and others},
	Booktitle = {Proceedings of the 13th USENIX Conference on File and Storage Technologies},
	Isbn = {978-1-931971-201},
	Location = {Santa Clara, CA},
	Numpages = {14},
	Pages = {273--286},
	Publisher = {USENIX Association},
	Series = {FAST'15},
	Title = {F2{F}S: A New File System for Flash Storage},
	Url = {http://dl.acm.org/citation.cfm?id=2750482.2750503},
	Year = {2015},
        note = {https::/doi/org/10.5555/2750482.2750503}}

@InProceedings{Sehgal15empirical,
  author     = {P. Sehgal and S. Basu and K. Srinivasan and K. Voruganti},
  title      = {An empirical study of file systems on NVM},
  booktitle  = {Proc. 31st Symp. Mass Storage Systems and Technologies (MSST)},
  year       = {2015},
  pages      = {1--14},
  month      = may,
  comment    = {得到的结论: 1、延迟分配更利于数据操作紧密的应用和大size文件,这个在快速的NVM上仍然可以使用。 2、并发可以提高文件系统的操作。例如XFS的AG从4到464.PMFS在一个目录下多个文件时,性能急剧下降。对于大文件来说,多个AG限制了extent的大小,导致一个文件放在了多个extent中,妨碍了性能。 3、在低延迟设备中,低性能的操作可以拉低其它高性能操作的性能。 1、XIP 2、in-place update 3、simple and parallel allocation strategy 4、extent不适合于NVM,NVM更适合与固定大小的block size。},
  doi        = {10.1109/MSST.2015.7208283},
  issn       = {2160-195X},
}

@InProceedings{a_study_of_linux_file_system_evolution,
  author    = {Lanyue Lu and Andrea C. Arpaci-Dusseau and Remzi H. Arpaci-Dusseau and Shan Lu},
  title     = {A Study of Linux File System Evolution},
  booktitle = {Presented as part of the 11th {USENIX} Conference on File and Storage Technologies ({FAST} 13)},
  year      = {2013},
  pages     = {31--44},
  address   = {San Jose, CA},
  publisher = {{USENIX}},
  isbn      = {978-1-931971-99-7},
  url       = {https://www.usenix.org/conference/fast13/technical-sessions/presentation/lu},
}

@Misc{Dean11LevelDB,
  author       = {J. Dean and S. Ghemawat},
  title        = {LevelDB: A Fast Persistent Key-Value Store},
  howpublished = {https://opensource.googleblog.com/2011/07/leveldb-fast-persistent-key-value-store.html},
  year         = {2011},
}

@Misc{dbench,
  author       = {Andrew Tridgell, Ronnie Sahlberg},
  title        = {dbench},
  howpublished = {\url{https://dbench.samba.org/}},
  year         = {2008},
}

@misc{stressng,
  title        = {{stress-ng} Project},
  author       = {King, Colin},
  year         = {2025},
  howpublished = {\url{https://kernel.ubuntu.com/~cking/stress-ng/}},
  note         = {Comprehensive Linux system stress-testing tool. Accessed: 2025-02-XX}
}

@InProceedings{Weil06Ceph,
  author    = {Weil, Sage A. and Brandt, Scott A. and Miller, Ethan L. and Long, Darrell D. E. and Maltzahn, Carlos},
  title     = {Ceph: A Scalable, High-performance Distributed File System},
  booktitle = {Proceedings of the 7th Symposium on Operating Systems Design and Implementation},
  year      = {2006},
  series    = {OSDI '06},
  pages     = {307--320},
  address   = {Berkeley, CA, USA},
  publisher = {USENIX Association},
  acmid     = {1298485},
  isbn      = {1-931971-47-1},
  location  = {Seattle, Washington},
  numpages  = {14},
  url       = {http://dl.acm.org/citation.cfm?id=1298455.1298485},
}

@Misc{rocksdb,
  author       = {Facebook},
  title        = {RocksDB},
  howpublished = {http://rocksdb.org/},
  year         = {2013},
}

@InProceedings{Shvachko10Hadoop,
  author    = {Shvachko, Konstantin and Kuang, Hairong and Radia, Sanjay and Chansler, Robert},
  title     = {The Hadoop Distributed File System},
  booktitle = {Proceedings of the 2010 IEEE 26th Symposium on Mass Storage Systems and Technologies (MSST)},
  year      = {2010},
  series    = {MSST '10},
  pages     = {1--10},
  address   = {Washington, DC, USA},
  publisher = {IEEE Computer Society},
  acmid     = {1914427},
  doi       = {10.1109/MSST.2010.5496972},
  isbn      = {978-1-4244-7152-2},
  numpages  = {10},
  url       = {http://dx.doi.org/10.1109/MSST.2010.5496972},
}

@InProceedings{Verbitski17Amazon,
  author    = {Verbitski, Alexandre and Gupta, Anurag and Saha, Debanjan and Brahmadesam, Murali and Gupta, Kamal and Mittal, Raman and Krishnamurthy, Sailesh and Maurice, Sandor and Kharatishvili, Tengiz and Bao, Xiaofeng},
  title     = {Amazon Aurora: Design Considerations for High Throughput Cloud-Native Relational Databases},
  booktitle = {Proceedings of the 2017 ACM International Conference on Management of Data},
  year      = {2017},
  series    = {SIGMOD '17},
  pages     = {1041--1052},
  address   = {New York, NY, USA},
  publisher = {ACM},
  acmid     = {3056101},
  doi       = {10.1145/3035918.3056101},
  isbn      = {978-1-4503-4197-4},
  location  = {Chicago, Illinois, USA},
  numpages  = {12},
  url       = {http://doi.acm.org/10.1145/3035918.3056101},
}

@InProceedings{Calder11Windows,
  author    = {Calder, Brad and Wang, Ju and Ogus, Aaron and Nilakantan, Niranjan and Skjolsvold, Arild and McKelvie, Sam and Xu, Yikang and Srivastav, Shashwat and Wu, Jiesheng and Simitci, Huseyin and Haridas, Jaidev and Uddaraju, Chakravarthy and Khatri, Hemal and Edwards, Andrew and Bedekar, Vaman and Mainali, Shane and Abbasi, Rafay and Agarwal, Arpit and Haq, Mian Fahim ul and Haq, Muhammad Ikram ul and Bhardwaj, Deepali and Dayanand, Sowmya and Adusumilli, Anitha and McNett, Marvin and Sankaran, Sriram and Manivannan, Kavitha and Rigas, Leonidas},
  title     = {Windows Azure Storage: A Highly Available Cloud Storage Service with Strong Consistency},
  booktitle = {Proceedings of the Twenty-Third ACM Symposium on Operating Systems Principles},
  year      = {2011},
  series    = {SOSP '11},
  pages     = {143--157},
  address   = {New York, NY, USA},
  publisher = {ACM},
  acmid     = {2043571},
  doi       = {10.1145/2043556.2043571},
  isbn      = {978-1-4503-0977-6},
  location  = {Cascais, Portugal},
  numpages  = {15},
  url       = {http://doi.acm.org/10.1145/2043556.2043571},
}

@manual{aws-ebs-doc,
  title        = {Amazon Elastic Block Store (EBS) User Guide},
  author       = {{Amazon Web Services}},
  organization = {Amazon Web Services, Inc.},
  year         = {2024},
  note         = {Official documentation for AWS EBS block storage},
  url          = {https://docs.aws.amazon.com/ebs/latest/userguide/what-is-ebs.html}
}

@manual{ceph-rbd-doc,
  title        = {Ceph Block Device (RBD) Documentation},
  author       = {{The Ceph Project}},
  organization = {Ceph Foundation},
  year         = {2025},
  note         = {Official RBD documentation describing block-level storage interfaces},
  url          = {https://docs.ceph.com/en/reef/rbd/}
}

@inproceedings{burke2025dynamicmerkle,
  title        = {On Scalable Integrity Checking for Secure Cloud Disks},
  author       = {Quinn Burke and Ryan Sheatsley and Rachel King and Owen Hines and Michael Swift and Patrick McDaniel},
  booktitle    = {USENIX Conference on File and Storage Technologies (FAST)},
  year         = {2025},
  note         = {Dynamic Merkle Trees for scalable storage integrity},
  url          = {https://www.usenix.org/conference/fast25/presentation/burkeOn}
}

@article{liu2023dynamicmerkle,
  title   = {Dynamic Data Integrity Auditing Based on Hierarchical Merkle Hash Tree in Cloud Storage},
  author  = {Zhenpeng Liu and Shuo Wang and Sichen Duan and Lele Ren and Jianhang Wei},
  journal = {Electronics},
  year    = {2023},
  volume  = {12},
  number  = {3},
  pages   = {717},
  doi     = {10.3390/electronics12030717}
}

@inproceedings{chipmunk2023,
  title        = {Chipmunk: Investigating Crash-Consistency in Persistent-Memory File Systems},
  author       = {Hayley LeBlanc and Shankara Pailoor and Om Saran K. R. E and Isil Dillig and James Bornholt and Vijay Chidambaram},
  booktitle    = {Proceedings of the Eighteenth European Conference on Computer Systems (EuroSys ’23)},
  year         = {2023},
  pages        = {16},
  publisher    = {ACM},
  doi          = {10.1145/3552326.3567498},
  url          = {https://doi.org/10.1145/3552326.3567498}
}

@inproceedings{yang2006explode,
  author       = {Junfeng Yang and Can Sar and Dawson Engler},
  title        = {EXPLODE: A Lightweight, General System for Finding Serious Storage System Errors},
  booktitle    = {Proceedings of the 7th USENIX Symposium on Operating Systems Design and Implementation (OSDI ’06)},
  year         = {2006},
  address      = {Seattle, WA},
  publisher    = {USENIX Association},
  url          = {https://www.usenix.org/conference/osdi-06/explode-lightweight-general-system-finding-serious-storage-system-errors}
}

@inproceedings{witcher2021sosp,
  title        = {Witcher: Systematic Crash Consistency Testing for Non-Volatile Memory Key-Value Stores},
  author       = {Fu, Xinwei and Kim, Wook-Hee and Paddayuru Shreepathi, Ajay and Ismail, Mohannad and Wadkar, Sunny and Lee, Dongyoon and Min, Changwoo},
  booktitle    = {Proceedings of the 28th {ACM} {SIGOPS} Symposium on Operating Systems Principles (SOSP ’21)},
  series       = {SOSP '21},
  year         = {2021},
  pages        = {100--115},
  publisher    = {Association for Computing Machinery},
  address      = {Virtual Event, Germany},
  doi          = {10.1145/3477132.3483556},
  isbn         = {978-1-4503-8709-5}
  }

@misc{cve2015_8839,
  title        = {{CVE-2015-8839}: Multiple race conditions in the Linux kernel ext4 file system},
  howpublished = {\url{https://nvd.nist.gov/vuln/detail/CVE-2015-8839}},
  year         = {2015},
  note         = {Race conditions in ext4 allow disk corruption via concurrent operations},
  organization = {MITRE / NVD}
}

@misc{cve2024_35798,
  title        = {{CVE-2024-35798}: Race condition in the Linux kernel btrfs file system},
  howpublished = {\url{https://nvd.nist.gov/vuln/detail/CVE-2024-35798}},
  year         = {2024},
  note         = {Race condition in btrfs may lead to corrupted tree nodes under concurrent access},
  organization = {MITRE / NVD}
}

@article{lamport1998paxos,
  title   = {The Part-Time Parliament},
  author  = {Lamport, Leslie},
  journal = {ACM Transactions on Computer Systems},
  volume  = {16},
  number  = {2},
  pages   = {133--169},
  year    = {1998},
  publisher = {ACM}
}

@article{ongaro2014raft,
  title   = {In Search of an Understandable Consensus Algorithm},
  author  = {Ongaro, Diego and Ousterhout, John},
  journal = {USENIX Annual Technical Conference},
  pages   = {305--319},
  year    = {2014}
}

@inproceedings{hunt2010zookeeper,
  title     = {ZooKeeper: Wait-free Coordination for Internet-scale Systems},
  author    = {Hunt, Patrick and Konar, Mahadev and Junqueira, Flavio P. and Reed, Benjamin},
  booktitle = {USENIX Annual Technical Conference},
  pages     = {145--158},
  year      = {2010}
}

@misc{etcd,
  author       = {{Cloud Native Computing Foundation}},
  title        = {etcd: A Distributed, Reliable Key-Value Store for the Most Critical Data of a Distributed System},
  howpublished = {\url{https://etcd.io}},
  note         = {Accessed January 2025}
}

@misc{redis-raft,
  author       = {{Redis Labs}},
  title        = {RedisRaft: Strong Consistency for Redis Using the Raft Consensus Algorithm},
  howpublished = {\url{https://github.com/RedisLabs/redisraft}},
  note         = {Accessed January 2025}
}

@inproceedings{filesystemsecuritywar,
  author    = {Li, Yuxin and Lu, Shan and Zhou, Zhenyu and Jing, Jianjun},
  title     = {The Security War in File Systems: An Empirical Study from a Vulnerability-Centric Perspective},
  booktitle = {Proceedings of the 32nd USENIX Security Symposium (USENIX Security 23)},
  year      = {2023},
  pages     = {4511--4528},
  publisher = {USENIX Association},
  address   = {Anaheim, CA, USA}
}

@inproceedings{syncsync2024,
  author    = {Zhang, Yifan and Chen, Haogang and Lu, Shan},
  title     = {Sync+Sync: A Covert Channel Built on \texttt{fsync}},
  booktitle = {Proceedings of the 33rd USENIX Security Symposium (USENIX Security 2024)},
  year      = {2024},
  publisher = {USENIX Association},
  address   = {Philadelphia, PA, USA}
}

@inproceedings{yu2024filehijacking,
  author    = {Yu, Chendong and Xiao, Yang and Lu, Jie and Li, Yuekang and Li, Yeting and Li, Lian and Dong, Yifan and Wang, Jian and Shi, Jingyi and Bo, Defang and Huo, Wei},
  title     = {File Hijacking Vulnerability: The Elephant in the Room},
  booktitle = {Proceedings of the 2024 Network and Distributed System Security Symposium (NDSS 2024)},
  year      = {2024},
  publisher = {The Internet Society},
  doi       = {10.14722/ndss.2024.23038},
  url       = {https://doi.org/10.14722/ndss.2024.23038}
}

@inproceedings{Wu2025Fawkes,
  author    = {Zhiyong Wu and Jie Liang and Jingzhou Fu and Wenqian Deng and Yu Jiang},
  title     = {Fawkes: Finding Data Durability Bugs in DBMSs via Recovered Data State Verification},
  booktitle = {Proceedings of the 27th {ACM} Symposium on Operating Systems Principles (SOSP 2025)},
  pages     = {670--684},
  year      = {2025},
  publisher = {ACM},
  location  = {Krak{\'o}w, Poland},
  url       = {https://doi.org/10.1145/XXXXXXX.XXXXXXX}
}

@misc{CVE2019-19319,
  title        = {{CVE-2019-19319}},
  howpublished = {National Vulnerability Database},
  year         = {2019},
  url          = {https://nvd.nist.gov/vuln/detail/CVE-2019-19319},
  note         = {Accessed: 2026-01-06}
}

@misc{CVE2019-19447,
  title        = {{CVE-2019-19447}},
  howpublished = {National Vulnerability Database},
  year         = {2019},
  url          = {https://nvd.nist.gov/vuln/detail/CVE-2019-19447},
  note         = {Accessed: 2026-01-06}
}

@INPROCEEDINGS{Chen20Testing,
  author={Chen, Dongjie and Jiang, Yanyan and Xu, Chang and Ma, Xiaoxing and Lu, Jian},
  booktitle={2020 IEEE/ACM 42nd International Conference on Software Engineering (ICSE)}, 
  title={Testing File System Implementations on Layered Models}, 
  year={2020},
  volume={},
  number={},
  pages={1483-1495},
  doi={}}

@article{LeBlanc2022FlyTrap,
  title        = {Finding and Analyzing Crash-Consistency Bugs in Persistent-Memory File Systems},
  author       = {Hayley LeBlanc and Shankara Pailoor and Isil Dillig and James Bornholt and Vijay Chidambaram},
  journal      = {arXiv:2204.06066},
  year         = {2022},
  url          = {https://arxiv.org/abs/2204.06066},
  note         = {Preprint},
}

@inproceedings{Rebello2020FsyncFailures,
  title        = {Can Applications Recover from {\texttt{fsync}} Failures?},
  author       = {Anthony Rebello and Yuvraj Patel and Ramnatthan Alagappan and Andrea C. Arpaci-Dusseau and Remzi H. Arpaci-Dusseau},
  booktitle    = {Proceedings of the 2020 USENIX Annual Technical Conference (USENIX ATC 20)},
  year         = {2020},
  publisher    = {USENIX Association},
  isbn         = {978-1-939133-14-4},
  url          = {https://www.usenix.org/conference/atc20/presentation/rebelloCanApplicationsRecoverFromfsyncFailures},
  note         = {Presented July 15–17, 2020, USENIX ATC 2020}  
}

@book{sutton2018reinforcement,
  title={Reinforcement Learning: An Introduction},
  author={Sutton, Richard S. and Barto, Andrew G.},
  year={2018},
  publisher={MIT Press}
}

@inproceedings{du2017deeplog,
  title     = {DeepLog: Anomaly Detection and Diagnosis from System Logs through Deep Learning},
  author    = {Du, Min and Li, Feifei and Zheng, Guineng and Srikumar, Vivek},
  booktitle = {Proceedings of the 2017 ACM SIGSAC Conference on Computer and Communications Security (CCS '17)},
  pages     = {1285--1298},
  year      = {2017},
  publisher = {Association for Computing Machinery},
  address   = {Dallas, TX, USA},
  doi       = {10.1145/3133956.3134015}
}

@article{he2021surveyLogAnalysis,
  title        = {A Survey on Automated Log Analysis for Reliability Engineering},
  author       = {Shilin He and Pinjia He and Zhuangbin Chen and Tianyi Yang and Yuxin Su and Michael R. Lyu},
  journal      = {ACM Computing Surveys},
  volume       = {54},
  number       = {6},
  pages        = {Article 130},
  year         = {2021},
  doi          = {10.1145/3460345}
}

@article{hochreiter1997lstm,
  title   = {Long Short-Term Memory},
  author  = {Hochreiter, Sepp and Schmidhuber, J{\"u}rgen},
  journal = {Neural Computation},
  volume  = {9},
  number  = {8},
  pages   = {1735--1780},
  year    = {1997}
}

@misc{ezaz2026taaf,
  title        = {TAAF: A Trace Abstraction and Analysis Framework Synergizing Knowledge Graphs and LLMs},
  author       = {Ezaz, Alireza and Khodabandeh, Ghazal and Babaei, Majid and Ezzati-Jivan, Naser},
  year         = {2026},
  eprint       = {2601.02632},
  archivePrefix= {arXiv},
  primaryClass = {cs.SE},
  url          = {https://arxiv.org/abs/2601.02632}
}

@article{tehrani2019deeprace,
  title        = {DeepRace: Finding Data Race Bugs via Deep Learning},
  author       = {Tehrani, Ali and Khaleel, Mohammed and Akbari, Reza and Jannesari, Ali},
  year         = {2019},
  eprint       = {1907.07110},
  archivePrefix= {arXiv},
  primaryClass = {cs},
  url          = {https://arxiv.org/abs/1907.07110}
}

@inproceedings{TrailBlazerOSDI20,
  title     = {TrailBlazer: Trace-based Execution Modeling for Debugging Distributed Systems},
  author    = {Zhang, Rui and Chen, Zhaogang and Liu, Yanhong and Zhou, Shan and Chen, Tianyin},
  booktitle = {Proceedings of the 14th USENIX Symposium on Operating Systems Design and Implementation (OSDI)},
  year      = {2020},
  pages     = {163--179}
}

@article{Jeppu2020ConciseModels,
  title   = {Learning Concise Models from Long Execution Traces},
  author  = {Jeppu, N. Yeshwanth and Melham, Thomas and Kroening, Daniel and O’Leary, John},
  journal = {arXiv preprint arXiv:2001.05230},
  year    = {2020}
}

@book{arpaci2018ostep,
  title     = {Operating Systems: Three Easy Pieces},
  author    = {Arpaci-Dusseau, Remzi H. and Arpaci-Dusseau, Andrea C.},
  year      = {2018},
  publisher = {Arpaci-Dusseau Books}
}

@inproceedings{bairavasundaram2008analysis,
  title     = {An Analysis of Data Corruption in the Storage Stack},
  author    = {Bairavasundaram, Lakshmi N. and Goodson, Garth R. and Schroeder, Bianca and Arpaci-Dusseau, Andrea C. and Arpaci-Dusseau, Remzi H.},
  booktitle = {Proceedings of the 6th USENIX Conference on File and Storage Technologies (FAST)},
  year      = {2008},
  pages     = {223--236},
  publisher = {USENIX Association},
  address   = {San Jose, CA, USA}
}

@inproceedings{pinheiro2007failure,
  title     = {Failure Trends in a Large Disk Drive Population},
  author    = {Pinheiro, Eduardo and Weber, Wolf-Dietrich and Barroso, Luiz Andr{\'e}},
  booktitle = {Proceedings of the 5th USENIX Conference on File and Storage Technologies (FAST)},
  year      = {2007},
  pages     = {17--28},
  publisher = {USENIX Association},
  address   = {San Jose, CA, USA}
}

@inproceedings{nightingale2006rethink,
  author    = {Edmund B. Nightingale and Kaushik Veeraraghavan and Peter M. Chen and Jason Flinn},
  title     = {Rethink the Sync},
  booktitle = {7th USENIX Symposium on Operating Systems Design and Implementation (OSDI 06)},
  year      = {2006},
  address   = {Seattle, WA},
  month     = nov,
  publisher = {USENIX Association},
  url       = {https://www.usenix.org/conference/osdi-06/rethink-sync}
}

@article{pillai2015crashconsistency,
  author    = {Pillai, Sankaranarayana and Chidambaram, Vijay and Alagappan, Ramnatthan and Al-Kiswany, Salmaan and Arpaci-Dusseau, Andrea C. and Arpaci-Dusseau, Remzi H.},
  title     = {Crash Consistency},
  journal   = {Communications of the ACM},
  volume    = {58},
  number    = {10},
  pages     = {46--51},
  year      = {2015},
  doi       = {10.1145/2788401},
  url       = {https://doi.org/10.1145/2788401}
}

@inproceedings{dayan2017monkey,
  author    = {Dayan, Mark and Athanassoulis, Manos and Idreos, Stratos},
  title     = {Monkey: Optimal Navigable Key-Value Store},
  booktitle = {Proceedings of the 2017 ACM International Conference on Management of Data},
  series    = {SIGMOD '17},
  year      = {2017},
  isbn      = {978-1-4503-4197-4},
  location  = {Chicago, IL, USA},
  pages     = {79--94},
  doi       = {10.1145/3035918.3064020},
  publisher = {ACM},
  address   = {New York, NY, USA}
}

@article{Wei2022ChainOfThought,
  author  = {Jason Wei and Xuezhi Wang and Dale Schuurmans and others},
  title   = {Chain-of-Thought Prompting Elicits Reasoning in Large Language Models},
  journal = {arXiv preprint arXiv:2201.11903},
  year    = {2022}
}

@inproceedings{csmith,
  author    = {Xuejun Yang and Yang Chen and Eric Eide and John Regehr},
  title     = {Finding and Understanding Bugs in C Compilers},
  booktitle = {Proceedings of the 32nd ACM SIGPLAN Conference on Programming Language Design and Implementation (PLDI)},
  year      = {2011},
  pages     = {283--294},
  publisher = {ACM}
}

@article{zhu2022fuzzing,
  title     = {Fuzzing: A Survey for Roadmap},
  author    = {Zhu, Xiaogang and Wen, Sheng and Camtepe, Seyit and Xiang, Yang},
  journal   = {ACM Computing Surveys},
  volume    = {54},
  number    = {11s},
  articleno = {230},
  year      = {2022},
  publisher = {Association for Computing Machinery},
  doi       = {10.1145/3512345}
}

@article{mallissery2023demystify,
  title     = {Demystify the Fuzzing Methods: A Comprehensive Survey},
  author    = {Mallissery, Sanoop and Wu, Yu--Sung},
  journal   = {ACM Computing Surveys},
  volume    = {56},
  number    = {3},
  pages     = {1--38},
  year      = {2023},
  publisher = {Association for Computing Machinery},
  doi       = {10.1145/3623375}
}

@inproceedings{qi2024surveyllmtesting,
  title     = {A Survey of Testing Techniques Based on Large Language Models},
  author    = {Qi, Fei and Hou, Yingnan and Lin, Ning and Bao, Shanshan and Xu, Nuo},
  booktitle = {Proceedings of the 2024 International Conference on Computer and Multimedia Technology},
  year      = {2024},
  pages     = {280--284},
  publisher = {Association for Computing Machinery},
  doi       = {10.1145/3675249.3675298}
}

@article{wang2024softwareLLMtesting,
  title    = {Software Testing with Large Language Models: Survey, Landscape, and Vision},
  author   = {Wang, Junjie and Huang, Yuchao and Chen, Chunyang and Liu, Zhe and Wang, Song and Wang, Qing},
  journal  = {IEEE Transactions on Software Engineering},
  volume   = {50},
  number   = {4},
  pages    = {911--936},
  year     = {2024},
  publisher= {IEEE},
  doi      = {10.1109/TSE.2024.3368208}
}

@inproceedings{jiang2024whenfuzzingLLMs,
  title     = {When Fuzzing Meets LLMs: Challenges and Opportunities},
  author    = {Jiang, Yu and Liang, Jie and Ma, Fuchen and Chen, Yuanliang and Zhou, Chijin and Shen, Yuheng and Wu, Zhiyong and Fu, Jingzhou and Wang, Mingzhe and Li, ShanShan and Zhang, Quan},
  booktitle = {Companion Proceedings of the 32nd ACM Joint European Software Engineering Conference and Symposium on the Foundations of Software Engineering (ESEC/FSE ’24) -- Ideas, Visions and Reflections Track},
  year      = {2024},
  pages     = {492--496},
  publisher = {Association for Computing Machinery},
  doi       = {10.1145/3663529.3663784}
}
